\RequirePackage{fix-cm}

\documentclass[%
a4paper, %
intlimits, %
twocolumn, %
]{revtex4-1}
\usepackage{cmap}

\usepackage{ucs}
\usepackage[utf8x]{inputenc}
\usepackage[T2A]{fontenc}
\usepackage[german,russian,english]{babel}

\usepackage{amsmath}
\usepackage{amssymb}

\usepackage{mathtools}
\mathtoolsset{
showonlyrefs,
mathic = true
}

\allowdisplaybreaks

\usepackage{nicefrac}

\usepackage{physymb}

\makeatletter
\def\ps@pprintTitle{%
     \let\@oddhead\@empty
     \let\@evenhead\@empty
     \let\@oddfoot\@empty
     \let\@evenfoot\@oddfoot}
\makeatother

\DeclareMathOperator{\Grad}{grad}
\DeclareMathOperator{\Div}{div}
\DeclareMathOperator{\Rot}{rot}

\renewcommand{\d}{\mathrm{d}}
\newcommand{\e}{\mathrm{e}}
\renewcommand{\i}{\mathrm{i}}

\newcommand{\const}{\mathrm{const}}

\DeclareMathOperator{\sign}{sign}

\newcommand{\crd}[1]{\underline{\vphantom{j}{#1}}}

\usepackage{tensor}

\begin{document}

\author{D. S. Kulyabov}
\email{yamadharma@gmail.com}
\author{A. V. Korolkova}
\email{avkorolkova@gmail.com}

\affiliation{Peoples' Friendship University of Russia}

\thanks{Published in: Kulyabov, D. S., Korolkova, A. V. \& Korolkov,
  V. I. Maxwell’s Equations in Arbitrary Coordinate System. Bulletin
  of Peoples’ Friendship University of Russia. Series
  «Mathematics. Information Sciences. Physics» 96--106 (2012).}

\title{Maxwell's Equations in Arbitrary Coordinate System}

\begin{abstract}
  The article is devoted to application of tensorial formalism for
  derivation of different types of Maxwell's equations. The Maxwell's
  equations are written in the covariant coordinate-free and the
  covariant coordinate forms.  Also the relation between vectorial and
  tensorial formalisms and differential operators for arbitrary
  holonomic coordinate system in coordinate form is given. The results
  obtained by tensorial and vectorial formalisms are verified in
  cylindrical and spherical coordinate systems.
\end{abstract}

\maketitle

\section{Introduction}

Problems of waveguide mathematical modelling sometimes need
curvilinear coordinate system to be applied.
The choice of specific
coordinate system is defined by the cross-section of the waveguide. 

Usually the description of waveguide model is based on Maxwell's
equations in Cartesian coordinate system.  With the help of vector
transformation property Maxwell's equations are rearranged for sertain
coordinate system (spherical or cylindrical). But in some problems,
e.g. simulation of a heavy-particle accelerator, the waveguide may
has the form of a cone or a hyperboloid. Another example of a
waveguide with a complex form is the Luneberg lens, which has the form
of a part of a sphere or a cylinder attached to a planar
waveguide. Therefore in the case of a waveguide with a complex form
the Maxwell's equations should be written in a arbitrary curvilinear
coordinate system.

It's well established to apply vectorial formalism to Maxwell's
equations. But in this case Maxwell's equations in a curvilinear
coordinate system are lengthy. In~\cite{PFUR-2011-2-kul::en} some
preliminary work on tensorial formalism resulting in a more compact form
of Maxwell's equations is made. The tensorial formalism has a
mathematical apparatus which allows to use covariant coordinate-free
form of Maxwell's equations. In this case the transition to a certain
coordinate system may be done on the final stage of research writing
down the results. But tensorial formalism can't be directly applied to
Maxwell's equations because the relation between vectorial and
tensorial formalisms should be proven before.

Different forms of Maxwell's equations are used in problems of finding
 Hamiltonian of
electromagnetic field    applied in variational
integrator (particularly, symplectic integrator) construction. The
main task is the fulfillment of the condition of symplectic structure
conservation during equations discretization. The several forms of
Maxwell's equations are used in electromagnetic field Hamiltonian
derivation:
\begin{itemize}
\item 3-vectors;
\item momentum representation (complex form is used);
\item momentum representation.
\end{itemize}

As a summary the main goals of the article may be formulated: to show
connection between vectorial and tensorial formalisms (section
\ref{s.1}); to apply tensorial formalism for different forms of
Maxwell's equations (section \ref{sec:maxwell_represent}); to verify
the obtained results by representing Maxwell's equations in spherical
and cylindrical coordinate systems (section \ref{sec:realisation}).

\section{Connection Between Vectorial and Tenzorial Formalisms}
\label{s.1}

Let's use the abstract indices formalism introduced
in~\cite{penrose-rindler-1987::en} in application to tensor algebra. In
\cite{penrose-rindler-1987::en} $\alpha$ is the abstract index,
$\crd{\alpha}$ --- a tensor component index. The usage of a component
index in some expression means that some arbitrary basis is introduced
in this equation and indices  obey the Einstein rule of summation
(the sum is taken over every numeric index which occurs in one term of
the expression twice --- top and bottom). Abstract indices have an
organizing value.

Let's consider an arbitrary $n$-dimensional vector and  space $V^{i}$
conjugated to $V^{i}$ space $V_{i}$.

In tensorial formalism the basis is given in coordinate form:
\[
\delta_{\crd{i}}^{i}=\frac{\partial}{\partial x^{\crd{i}}} \in V^{i},\quad 
\delta^{\crd{i}}_{i}=\d x^{\crd{i}} \in V_{i},\quad \crd{i}=\overline{1,n}.
\]
In vectorial formalism the basis is given by elements with the length
$\d s^{\crd{i}'}$ upon the corresponding coordinate:
\[
\delta_{\crd{i}'}^{i}=\frac{\partial}{\partial s^{\crd{i}'}},\quad 
\delta^{\crd{i}'}_{i}=\d s^{\crd{i}'},\quad \crd{i}'=\overline{1,n}.
\]

Nonholonomic basis usually provides some comfort.
If this:
\begin {itemize}
\item retention values ​​coordinate conversion
  (i.e., moving distance in distance, angles in the angles and so on);
\item indistinguishability of contravariant and covariant vectors that
  allows to use only one index type.
\end{itemize}

In tensor form:
\begin{equation}
\label{eq:ds2-t}
\d s^2= g_{\crd{i}\crd{j}}\d x^{\crd{i}}\d x^{\crd{j}},\quad
\crd{i},\,\crd{j}=\overline{1,n},
\end{equation}
where $g_{\crd{i}\crd{j}}$~--- metric tensor.

In vector form:
\begin{equation}
\label{eq:ds2-v}
\d s^2=g_{\crd{i}' \crd{j}'}\d s^{\crd{i}'}\d s^{\crd{j}'},
\quad \crd{i}',\, \crd{j}'=\overline{1,n}.
\end{equation}
In the case of orthogonal basis, 
\eqref{eq:ds2-v} has the form of:
\begin{equation}
\label{eq:ds2-v2}
\d s^2=g_{\crd{i}'\crd{i}'}\d s^{\crd{i}'}\d s^{\crd{i}'}, \quad
\crd{i}'=\overline{1,n}. 
\end{equation}

let's express the vector basis through the tensor one:
\[
\d s^{\crd{i}'}=h_{\crd{i}}^{\crd{i}'}\d x^{\crd{i}},\quad 
\frac{\partial}{\partial
  s^{\crd{i}'}}=h_{\crd{i}'}^{\crd{i}}\frac{\partial}{\partial
  x^{\crd{i}}}. 
\]
Where $h_{\crd{i}}^{\crd{i}'}$, $h_{\crd{i}'}^{\crd{i}}$,
$\crd{i},\,\crd{i}'=\overline{1,n}$,~--- matrix of Jacobi.

For orthogonal basis from \eqref{eq:ds2-v2}
\[
g_{\crd{i}\crd{i}}\d x^{\crd{i}}\d
x^{\crd{i}}= g_{\crd{i}'\crd{i}'}
h_{\crd{i}}^{\crd{i}'}h_{\crd{i}}^{\crd{i}'}\d x^{\crd{i}}\d
x^{\crd{i}}, \quad \crd{i},\,\crd{i}'=\overline{1,n}.
\]

Let's introduce the notation (for orthogonal coordinate system):
\[
\left(h_{\crd{i}}\right)^{2}:= 
h_{\crd{i}}^{\crd{i}'}h_{\crd{i}}^{\crd{i}'} =
\frac{g_{\crd{i}\crd{i}}}{g_{\crd{i}'\crd{i}'}},\quad 
h_{\crd{i}} := h_{\crd{i}}^{\crd{i}'} =
\sqrt{\frac{g_{\crd{i}\crd{i}}}{g_{\crd{i}'\crd{i}'}}},
\quad
\crd{i},\,\crd{i}'=\overline{1,n}.
\]
Variables $h_{\crd{i}}$ are called Lame coefficients~\cite{mors::en}.

Let's express vector $f^{i}\in V^{i}$ by its components
$f^{\crd{i}}$ in tensor $\delta_{\crd{i}}^{i}$ and vector
$\delta_{\crd{i}'}^{i}$ basises:
\[
\begin{gathered}
f^{i}=f^{\crd{i}}\delta_{\crd{i}}^{i}=f^{\crd{i}}\frac{\partial} 
{\partial x^{\crd{i}}},\\
f^{i}=f^{\crd{i}'}\delta_{\crd{i}'}^{i}=f^{\crd{i}'}\frac{\partial}{\partial
s^{\crd{i}'}}=f^{\crd{i}'}\frac{1}{h_{\crd{i}}^{\crd{i}'}}\frac{\partial}{\partial
x^{\crd{i}}}.
\end{gathered}
\]
And then
\begin{equation}
\label{eq:vec_nonholonom}
f^{\crd{i}'}=f^{\crd{i}} h_{\crd{i}}^{\crd{i}'}, \quad \crd{i},\,\crd{i}'=\overline{1,n}.
\end{equation}

In the similar way, for covectors:
\[
\begin{gathered}
f_{i}=f_{\crd{i}}\delta^{\crd{i}}_{i}=f_{\crd{i}}\d x^{\crd{i}},\\
f_{i}=f_{\crd{i}'}\delta^{\crd{i}'}_{i}=f_{\crd{i}'}\d
s^{\crd{i}'}=f_{\crd{i}'}h_{\crd{i}}^{\crd{i}'}
\d x^{\crd{i}},
\end{gathered}
\]
and then 
\begin{equation}
\label{eq:covec_nonholonom}
f_{\crd{i}'}=f_{\crd{i}} \frac{1}{h_{\crd{i}}^{\crd{i}'}},\quad \crd{i},\,\crd{i}'=\overline{1,n}.
\end{equation}

So the connection between tensorial and vectorial formalisms is
proved.

\section{Tensorial Notation  of Differential Operators in Components}
\label{s.2}
Let's present the differential operators in the components (for connections
associated with metric).

The expression for gradient (the variable $\varphi$ is a scalar):
\begin{equation}
\label{eq:grad}
\begin{gathered}
(\Grad \varphi)_i= (\Grad \varphi)_{\crd{i}}
\delta_{i}^{\crd{i}},\\ 
(\Grad \varphi)_{\crd{i}}=\nabla_{\crd{i}}\varphi
=\partial_{\crd{i}}\varphi, \quad \crd{i}=\overline{1,n}.
\end{gathered}
\end{equation}

The expression for an arbitrary vector divergence $\vec{f}\in
V^{i}$ is:
\begin{multline}
  \label{eq:div}
\Div  \vec{f}= \nabla_i f^i= f_{, i}^i- \Gamma_{ji}^i f^j= f_{, i}^i-
f^i\frac{(\sqrt{|g|})_{, i}}{\sqrt{|g|}}= {} \\
{} = \frac{1}{\sqrt{|g|}}\partial_i\left(\sqrt{|g|} f^i\right),
\end{multline}
or in components:
\begin{equation}
  \label{eq:div_comp}
\Div  \vec{f}=\frac{1}{\sqrt{|g|}}\partial_{\crd{i}}\left(\sqrt{|g|}
  f^{\crd{i}}\right), \quad \crd{i}=\overline{1,n}.
\end{equation}
Variable $g$ is $\det\left(g_{\crd{i} \crd{i}}\right)$.

Because of the nonnegativity of radical expression and because of
$\mathbb{M}^4$ $g < 0$ in Minkowsky space let's use the following
notation $|g|$.

The expression for rotor is valid only in $\mathbb{E}^3$ space:
\begin{equation}
\label{eq:rot}
\begin{gathered}
\left(\Rot \vec{f}\right)^{i}=\left[\vec{\nabla}, \vec{f}\right]=
\left(\Rot \vec{f}\right)^{\crd{i}} \delta_{\crd{i}}^{i}, \\ 
\left(\Rot \vec{f}\right)^{\crd{i}}=e^{\crd{i}\crd{j}\crd{k}}
\nabla_{\crd{j}} f_{\crd{k}},\quad \crd{i},\, \crd{j},\,
\crd{k}=\overline{1,3},
\end{gathered}
\end{equation}
where $e^{\crd{i}\crd{j}\crd{k}}$ is the alternating tensor expressed by
Levi--Civita simbol $\varepsilon^{\crd{i}\crd{j}\crd{k}}$:
\[
e_{\crd{i}\crd{j}\crd{k}}=\sqrt{g^{(3)}}\varepsilon_{\crd{i}\crd{j}\crd{k}},\quad
e^{\crd{i}\crd{j}\crd{k}}=\frac{1}{\sqrt{g^{(3)}}}\varepsilon^{\crd{i}\crd{j}\crd{k}},
\quad \crd{i},\, \crd{j},\, \crd{k}=\overline{1,3}.
\]

In general
\begin{equation}
\label{eq:e_alter}
\begin{gathered}
  e_{\crd{a_1}\crd{a_2}\dots\crd{a_n}} = 
  \sqrt{|g^{(n)}|} \varepsilon_{\crd{a_1}\crd{a_2}\dots\crd{a_n}},\\
  e^{\crd{a_1}\crd{a_2}\dots\crd{a_n}} =
  \frac{\sqrt{|g^{(n)}|}}{g^{(n)}}
  \varepsilon^{\crd{a_1}\crd{a_2}\dots\crd{a_n}}
  = \frac{\sgn g^{(n)}}{\sqrt{|g^{(n)}|}}
  \varepsilon^{\crd{a_1}\crd{a_2}\dots\crd{a_n}}, \\
  \crd{a_1},\, \crd{a_2},\dots , \crd{a_n}=\overline{1,n}.
\end{gathered}
\end{equation}

From \eqref{eq:div} for divergence and \eqref{eq:grad} for gradient
one can get Laplacian:
\begin{multline}
  \label{eq:lapl}
\Delta \varphi= \nabla_i\left(\nabla^i
  \varphi\right)=\nabla_i\left(g^{ij}(\Grad \varphi)_j\right)= {} \\
{} = \nabla_i\left(g^{ij}\partial _j
  \varphi\right)=\frac{1}{\sqrt{|g|}}\partial_i\left(\sqrt{|g|}g^{ij}\partial_j
\varphi\right).
\end{multline}

\section{Maxwell's Equations Presentation}
\label{sec:maxwell_represent}

Let's consider Maxwell's equations in CGS-system:
\begin{equation}
  \label{eq:maxwell}
  \begin{gathered}
    \vec \nabla \times \vec E = - \frac{1}{c} \frac{\partial
      \vec{B}}{\partial t}; \\
    \vec{\nabla}\cdot\vec{D} = 4\pi\rho; \\
    \vec \nabla \times \vec H = \frac{1}{c} \frac{\partial
      \vec{D}}{\partial t} + \frac{4\pi}{c} \vec{j}; \\
    \vec{\nabla}\cdot\vec{B} = 0.
  \end{gathered}
\end{equation}
Here $\vec{E}$ and $\vec{H}$~--- electric and magnetic intensities,
$\vec{j}$ is the current density, $\rho$ is the charge density, $c$ is the
light velocity.

\subsection{Maxwell's Equations Covariant Form by 3-vectors}
\label{sec:3-vector}

Let's express the equation \eqref{eq:maxwell} in the covariant form
\begin{equation}
  \label{eq:3-vector:cov}
  \begin{gathered}
   e^{ijk} \nabla_{j} E_{k} = - \nabla_{0} B^{i}; \\
    \nabla_{i}D^{i} = 4\pi\rho; \\
    e^{ijk} \nabla_{j} H_{k} = \nabla_{0} D^i + \frac{4\pi}{c} j^{i}; \\
    \nabla_{i} B^{i} = 0.
  \end{gathered}
\end{equation}

Let's rewrite the expression \eqref{eq:maxwell} in the tensorial formalism
components with the help of \eqref{eq:rot} and~\eqref{eq:div_comp}:
\begin{equation}
  \label{eq:maxwell_comp}
  \begin{gathered}
    \frac{1}{\sqrt{g^{(3)}}}\left[\partial_{\crd{j}}E_{\crd{k}}-\partial_{\crd{k}}E_{\crd{j}}\right]
    = -\frac{1}{c}\partial_t B^{\crd{i}}, \quad \crd{i},\,\crd{j},\,
    \crd{k}=\overline{1,3},\\
    \frac{1}{\sqrt{g^{(3)}}}\partial_{\crd{i}}\left(\sqrt{g^{(3)}}
      D^{\crd{i}}\right)= 4\pi \rho, \quad \crd{i}=\overline{1,3},\\
    \frac{1}{\sqrt{g^{(3)}}}\left[\partial_{\crd{j}}H_{\crd{k}}-\partial_{\crd{k}}H_{\crd{j}}\right]
    = -\frac{1}{c}\partial_t D^{\crd{i}}+\frac{4\pi}{c}j^{\crd{i}}, \quad \crd{i},\,\crd{j},\,
    \crd{k}=\overline{1,3},\\
    \frac{1}{\sqrt{g^{(3)}}}\partial_{\crd{i}}\left(\sqrt{g^{(3)}}
      B^{\crd{i}}\right)= 0, \quad \crd{i}=\overline{1,3}.
  \end{gathered}
\end{equation}

\subsection{Maxwell's Equations Covariant Form by 4-vectors}
\label{sec:4-vector}

Let's rewrite \eqref{eq:maxwell} with the help of electromagnetic
field tensors $F_{\alpha\beta}$ and
$G_{{\alpha}{\beta}}$\cite{minkowski:1910::en}:
\begin{gather}
\nabla_{{\alpha}} F_{{\beta}{\gamma}}+ \nabla_{{\beta}}
F_{{\gamma}{\alpha}}+\nabla_{{\gamma}} F_{{\alpha}{\beta}} = 
F_{[\alpha \beta ; \gamma]} = 0,
\label{eq:m:tensor:2}
\\
\nabla_{{\alpha}} G^{{\alpha}{\beta}}=\frac{4 \pi}{c}j^{{\beta}},
\label{eq:m:tensor}
\end{gather}
where
\begin{gather}
F_{\crd{\alpha}\crd{\beta}}=
\begin{pmatrix} 
0 & {E}_1 & {E}_2 & {E}_3 \\ 
-{E}_1 & 0 & -{B}^3 & {B}^2 \\
-{E}_2 & {B}^3 & 0 & -{B}^1 \\ 
-{E}_3 & -{B}^2 & {B}^1 & 0 
\end{pmatrix},
\\
F^{\crd{\alpha}\crd{\beta}}=
\begin{pmatrix} 
0 & -{E}^1 & -{E}^2 & -{E}^3 \\ 
{E}^1 & 0 & -{B}_3 & {B}_2 \\
{E}^2 & {B}_3 & 0 & -{B}_1 \\ 
{E}^3 & -{B}_2 & {B}_1 & 0 
\end{pmatrix},
\\
G^{\crd{\alpha}\crd{\beta}}=
\begin{pmatrix} 
0 & -{D}^1 & -{D}^2 & -{D}^3 \\ 
{D}^1 & 0 & -{H}_3 & {H}_2 \\
{D}^2 & {H}_3 & 0 & -{H}_1 \\ 
{D}^3 & -{H}_2 & {H}_1 & 0 
\end{pmatrix},
\\
G_{\crd{\alpha}\crd{\beta}}=
\begin{pmatrix} 
0 & {D}_1 & {D}_2 & {D}_3 \\ 
-{D}_1 & 0 & -{H}^3 & {H}^2 \\
-{D}_2 & {H}^3 & 0 & -{H}^1 \\ 
-{D}_3 & -{H}^2 & {H}^1 & 0 
\end{pmatrix},
\end{gather}
  $E_{\crd{i}}$, $H^{\crd{i}}$, $\crd{i}=\overline{1,3}$,~---
  components of electric and magnetic fields intensity vectors;
  $D_{\crd{i}}$, $B^{\crd{i}}$, $\crd{i}=\overline{1,3}$,~---
  components of vectors of electric and magnetic
  induction\footnote{Note that namely $B^{i}$ has the physical meaning
    of the magnetic field.}.

The equation \eqref{eq:m:tensor:2} may be rewritten in a simpler form
\begin{equation}
  \label{eq:m:tensor:2:dual}
  \nabla_{\alpha} \prescript{*}{}{F}^{\alpha\beta} = 0.
\end{equation}
Where the tensor $\prescript{*}{}{F}^{\alpha\beta}$ dual conjugated to
${F}^{\alpha\beta}$ is introduced
\begin{equation}
  \label{eq:tilde_g}
  \prescript{*}{}{F}^{\alpha\beta} = \frac{1}{2} e^{\alpha\beta\gamma\delta} F_{\gamma\delta},
\end{equation}
where $e^{\alpha\beta\gamma\delta}$ is the alternating tensor (see~\eqref{eq:e_alter}):
\begin{equation}
  \label{eq:e4}
  e_{\crd{\alpha}\crd{\beta}\crd{\gamma}\crd{\delta}} = \sqrt{-g}\varepsilon_{\crd{\alpha}\crd{\beta}\crd{\gamma}\crd{\delta}},\quad
  e^{\crd{\alpha}\crd{\beta}\crd{\gamma}\crd{\delta}} = -\frac{1}{\sqrt{-g}}\varepsilon^{\crd{\alpha}\crd{\beta}\crd{\gamma}\crd{\delta}}.
\end{equation}

Similarly
\begin{equation}
  \label{eq:tilde_other}
  \begin{gathered}
    \prescript{*}{}{F}_{\alpha\beta} = \frac{1}{2}
    e_{\alpha\beta\gamma\delta} F^{\gamma\delta}, \\
    \prescript{*}{}{G}_{\alpha\beta} = \frac{1}{2}
    e_{\alpha\beta\gamma\delta} G^{\gamma\delta}, \\
    \prescript{*}{}{G}^{\alpha\beta} = \frac{1}{2}
    e^{\alpha\beta\gamma\delta} G_{\gamma\delta}.
  \end{gathered}
\end{equation}

We write in components:
\begin{equation}
\begin{gathered}
\prescript{*}{}{F}_{\crd{\alpha}\crd{\beta}}=
\sqrt{-g}
\begin{pmatrix} 
0 & {B}_1 & {B}_2 & {B}_3 \\ 
-{B}_1 & 0 & {E}^3 & -{E}^2 \\
-{B}_2 & -{E}^3 & 0 & {E}^1 \\ 
-{B}_3 & {E}^2 & -{E}^1 & 0 
\end{pmatrix},
\\
\prescript{*}{}{F}^{\crd{\alpha}\crd{\beta}}=
\frac{1}{\sqrt{-g}}
\begin{pmatrix} 
0 & -{B}^1 & -{B}^2 & -{B}^3 \\ 
{B}^1 & 0 & {E}_3 & -{E}_2 \\
{B}^2 & -{E}_3 & 0 & {E}_1 \\ 
{B}^3 & {E}_2 & -{E}_1 & 0 
\end{pmatrix},
\\
\prescript{*}{}{G}^{\crd{\alpha}\crd{\beta}}=
\frac{1}{\sqrt{-g}}
\begin{pmatrix} 
0 & -{H}^1 & -{H}^2 & -{H}^3 \\ 
{H}^1 & 0 & {D}_3 & -{D}_2 \\
{H}^2 & -{D}_3 & 0 & {D}_1 \\ 
{H}^3 & {D}_2 & -{D}_1 & 0 
\end{pmatrix},
\\
\prescript{*}{}{G}_{\crd{\alpha}\crd{\beta}}=
\sqrt{-g}
\begin{pmatrix} 
0 & {H}_1 & {H}_2 & {H}_3 \\ 
-{H}_1 & 0 & {D}^3 & -{D}^2 \\
-{H}_2 & -{D}^3 & 0 & {D}^1 \\ 
-{H}_3 & {D}^2 & -{D}^1 & 0 
\end{pmatrix},
\end{gathered}
\end{equation}

The ordered pair $(E_{\crd{i}}, B^{\crd{i}})$ ($F_{\crd{\alpha}\crd{\beta}} \sim (E_{\crd{i}},
B^{\crd{i}})$) may be assigned to $F_{\crd{\alpha}\crd{\beta}}$ by following
\begin{equation}
  \label{eq:F_equiv_para}
  F_{0\crd{i}} = E_{i}, \quad F_{\crd{i}\crd{j}} = - \varepsilon_{\crd{i}\crd{j}\crd{k}} B^{\crd{k}}.
\end{equation}

So the following expressions may be written
\begin{equation}
  \label{eq:all_equiv}
  \begin{gathered}
    F_{\crd{\alpha}\crd{\beta}} \sim (E_{\crd{i}}, B^{\crd{i}}),
    \qquad 
    F^{\crd{\alpha}\crd{\beta}} \sim (-E^{\crd{i}}, B_{\crd{i}}), \\
    G_{\crd{\alpha}\crd{\beta}} \sim (D_{\crd{i}}, H^{\crd{i}}), 
    \qquad 
    G^{\crd{\alpha}\crd{\beta}} \sim  (-D^{\crd{i}}, H_{\crd{i}}), \\
    \prescript{*}{}{F}_{\crd{\alpha}\crd{\beta}} \sim 
    \sqrt{-g} (B_{\crd{i}}, -E^{\crd{i}}), 
    \qquad 
    \prescript{*}{}{F}^{\crd{\alpha}\crd{\beta}} \sim 
    \frac{1}{\sqrt{-g}} (-B^{\crd{i}}, -E_{\crd{i}}), \\
    \prescript{*}{}{G}_{\crd{\alpha}\crd{\beta}} \sim 
    \sqrt{-g} (H_{\crd{i}}, -D^{\crd{i}}), 
    \qquad 
    \prescript{*}{}{G}^{\crd{\alpha}\crd{\beta}} \sim 
    \frac{1}{\sqrt{-g}} (-H^{\crd{i}}, -D_{\crd{i}}).
  \end{gathered}
\end{equation}

\subsection{Complex Form of Maxwell's Equations}
\label{sec:complex}

The complex form of Maxwell's equations was considered by various
authors \cite{stratton:1948::en}, \cite{silberstein:1907::en}.

Similar to \eqref{eq:all_equiv} let's introduce correspondence between
an ordered pair and a complex 3-vector
\begin{equation}
  \label{eq:complex:eqiv}
  \begin{gathered}
    F^{\crd{i}} \sim (E^{\crd{i}}, B^{\crd{i}}), 
    \quad 
    F^{\crd{i}} = E^{\crd{i}} + \i B^{\crd{i}}; \\
    G^{\crd{i}} \sim (D^{\crd{i}}, H^{\crd{i}}), 
    \quad 
    G^{\crd{i}} = D^{\crd{i}} + \i H^{\crd{i}}.
  \end{gathered}
\end{equation}

Let's express intensity and induction by means of complex vectors
\begin{equation}
  \label{eq:complex:vectors}
  \begin{gathered}
    E^{i} = \frac{F^{i} + \Bar{F}^{i}}{2}, \quad 
    B^{i} = \frac{F^{i} - \Bar{F}^{i}}{2 \i}, \\
    D^{i} = \frac{G^{i} + \Bar{G}^{i}}{2}, \quad 
    H^{i} = \frac{G^{i} - \Bar{G}^{i}}{2 \i}.
  \end{gathered}
\end{equation}

Two complementary vectors
\begin{equation}
  \label{eq:complex:K_L}
  K^{i} = \frac{G^{i} + F^{i}}{2}, \quad L^{i} = \frac{\Bar{G}^{i} - \Bar{F}^{i}}{2}.
\end{equation}

The expression \eqref{eq:3-vector:cov} assumes the form
\begin{equation}
  \label{eq:complex:maxwell_K_L}
  \begin{gathered}
    \nabla_{i} (K^{i} + L^{i}) = 4\pi\rho; \\
    -\i \nabla_{0} (K^{i} - L^{i}) + e^{ijk} \nabla_{j} (K_{k} -
    L_{k}) = \i \frac{4\pi}{c} j^{i}.
  \end{gathered}
\end{equation}

\subsubsection{Complex Form of Maxwell's Equations in Vacuum}
\label{sec:complex:vacuum}

From $D^{i} = E^{i}$, $H^{i} = B^{i}$ and \eqref{eq:complex:K_L}
it follows
\begin{equation}
  \label{eq:complex:K_L:vacuum}
  K^{i} = E^{i} + \i B^{i} = F^{i}, \quad L^{i} = 0.
\end{equation}

Then the equations \eqref{eq:complex:maxwell_K_L} will have the form
\begin{equation}
  \label{eq:complex:maxwell_K_L:vacuum}
  \begin{gathered}
    \nabla_{i} F^{i} = 4\pi\rho; \\
    -\i \nabla_{0} F^{i} + e^{ijk} \nabla_{j} F_{k} 
    = \i \frac{4\pi}{c} j^{i}.
  \end{gathered}
\end{equation}

\subsubsection{Complex Representation of Maxwell's Equations in
  Homogeneous Isotropic Space}
\label{sec:complex:isotrop}

In homogeneous isotropic space the following relations $D^{i} =
\varepsilon E^{i}$, $\mu H^{i} = B^{i}$ (where $\varepsilon$~---
dielectric permittivity and
$\mu$ --- magnetic permeability) 
 are correct.

The resulting expressions may be simplified as follows. In
\eqref{eq:complex:maxwell_K_L:vacuum} we need the formal substitutions $c \to
c' = \frac{c}{\sqrt{\varepsilon\mu}}$ (the speed of light in vacuum is
substituted by the speed of light in medium) and $j^{\alpha} \to
\frac{j^{\alpha}}{\sqrt{\varepsilon}}$.
The result:
\begin{equation}
  \label{eq:complex:maxwell:isotrop}
  \begin{gathered}
    F^{i} = \sqrt{\varepsilon} E^{i} + \i \frac{1}{\sqrt{\mu}} B^{i}, \\
    \nabla_{i} F^{i} = \frac{4\pi}{\sqrt{\varepsilon}}\rho; \\
    e^{ijk} \nabla_{j} F_{k} = \i \frac{4\pi\sqrt{\mu}}{c} j^{i}
    + \i \frac{\sqrt{\varepsilon\mu}}{c} \frac{\partial F^{i}}{\partial t}.
  \end{gathered}
\end{equation}

This representation of Maxwell's equations has several names.
In particular, it is known as a representation of the Riemann--Sielberstein.

\subsection{Momentum Representation of Maxwell's Equations}
\label{sec:impuls}

Let's expand the vectors of electric and magnetic fields intensity in
a wavevector Fourier series $k^j$, $j$~--- abstract index:
\begin{equation}
  \label{eq:m:fourie}
  \begin{gathered}
    \Hat{E}^i(t,k_j)=\frac{1}{\sqrt{(2\pi)^3}}\int \d^3 x^j \sqrt{g^{(3)}}
    E^i(t,x^j) \e^{-\i k_jx^j},\\
    \Hat{H}^i(t,k_j)=\frac{1}{\sqrt{(2\pi)^3}}\int \d^3 x^j \sqrt{g^{(3)}}
    H^i(t,x^j) \e^{-\i k_jx^j},\\
    \Hat{B}^i(t,k_j)=\frac{1}{\sqrt{(2\pi)^3}}\int \d^3 x^j \sqrt{g^{(3)}}
    B^i(t,x^j) \e^{-\i k_jx^j},\\
    \Hat{D}^i(t,k_j)=\frac{1}{\sqrt{(2\pi)^3}}\int \d^3 x^j \sqrt{g^{(3)}}
    D^i(t,x^j) \e^{-\i k_jx^j},\\
    \Hat{\rho}(t,k_j)=\frac{1}{\sqrt{(2\pi)^3}}\int \d^3 x^j \sqrt{g^{(3)}}
    \rho(t,x^j) \e^{-\i k_jx^j},\\
    \Hat{\jmath}^i(t,k_j)=\frac{1}{\sqrt{(2\pi)^3}}\int \d^3 x^j \sqrt{g^{(3)}}
    j^i(t,x^j) \e^{-\i k_jx^j}.
  \end{gathered}
\end{equation}
And inverse transvorm is
\begin{equation}
  \label{eq:m:fourie}
  \begin{gathered}
    E^i(t,x^j)=\frac{1}{\sqrt{(2\pi)^3}}\int \d^3 k_j \sqrt{\Hat{g}^{(3)}}
    \Hat{E}^i(t,k_j) \e^{\i k_jx^j},\\
    H^i(t,x^j)=\frac{1}{\sqrt{(2\pi)^3}}\int \d^3 k_j \sqrt{\Hat{g}^{(3)}}
    \Hat{H}^i(t,k_j) \e^{\i k_jx^j},\\
    B^i(t,x^j)=\frac{1}{\sqrt{(2\pi)^3}}\int \d^3 k_j \sqrt{\Hat{g}^{(3)}}
    \Hat{B}^i(t,k_j) \e^{\i k_jx^j},\\
    D^i(t,x^j)=\frac{1}{\sqrt{(2\pi)^3}}\int \d^3 k_j \sqrt{\Hat{g}^{(3)}}
    \Hat{D}^i(t,k_j) \e^{\i k_jx^j}, \\
    \rho(t,x^j)=\frac{1}{\sqrt{(2\pi)^3}}\int \d^3 k_j \sqrt{\Hat{g}^{(3)}}
    \Hat{\rho}(t,k_j) \e^{\i k_jx^j}, \\
    j^i(t,x^j)=\frac{1}{\sqrt{(2\pi)^3}}\int \d^3 k_j \sqrt{\Hat{g}^{(3)}}
    \Hat{\jmath}^i(t,k_j) \e^{\i k_jx^j}.
  \end{gathered}
\end{equation}

Let's note that the vector components $E^i(t,x^j)$ and $\Hat{E}^i(t,k_j)$,
(similarly: $H^i(t,x^j)$ and $\Hat{H}^i(t,k_j)$, $D^i(t,x^j)$ and $\Hat{D}^i(t,k_j)$,
$B^i(t,x^j)$ and $\Hat{B}^i(t,k_j)$, $j^i(t,x^j)$ and $\Hat{\jmath}^i(t,k_j)$) are used in
different basises:
\begin{equation}
\begin{gathered}
E^i(t,x^j)=E^{\crd{i}}(t,x^j)\delta^i_{\crd{i}},\\
\Hat{E}^i(t,k_j)=\Hat{E}^{\crd{\Hat{\imath}}}(t,k_j)\delta^i_{\crd{\Hat{\imath}}}, \\
\Hat{g}^{(3)} := \det g_{\crd{\Hat{\imath}} \crd{\Hat{\jmath}}}, \quad 
\d s^2 = g_{\crd{\Hat{\imath}} \crd{\Hat{\jmath}}} \d x^{\crd{\Hat{\imath}}} \d x^{\crd{\Hat{\jmath}}}.
\end{gathered}
\end{equation}
where the basis $\delta^i_{\Hat{\crd{\imath}}}$ is given according to the
vector $k_i$.  For all $k_i$ the independent basis is defined. 
We can use expressions under integral sign putting down
Maxwell's equations from \eqref{eq:m:fourie}. Or
use the formulas for the Fourier transforms.
\begin{equation}
\label{eq:fourier_transforms}
\begin{gathered}
\widehat{(af(x^i)+bg(x^i))} = a \Hat{f}(k_i) + b \Hat{g}(k_i), \quad a,
b = \const, \\
\widehat{\frac{\partial f(x^i) }{\partial x^{\crd{j}}}} = \i
k_{\crd{j}} \Hat{f}(k^i), \\
\widehat{f(x^i) g(x^i)} = \frac{1}{\sqrt{(2\pi)^3}} (\Hat{f} *
\Hat{g}) (k_{i}), \\ 
\shortintertext{where} 
(\Hat{f} *\Hat{g}) (k_{i}) =
\int_{-\infty}^{\infty} \Hat{f}(k_{i} - s_{i}) \Hat{g}(s_{i}) \d^3 s_{i}.
\end{gathered}
\end{equation}

In the simplest case ($g = \const$) 

\begin{equation}
  \label{eq:maxwell:k}
  \begin{gathered}
    \i \frac{1}{\sqrt{g^{(3)}}}\varepsilon^{ijk} k_j E_{k}(t,k_j)=-\frac{1}{c}\partial_t B^i(t, k_j),\\
    \i\frac{1}{\sqrt{g^{(3)}}}\varepsilon^{ijk}k_j H_k(t,k_j)=\frac{1}{c}\partial_t D^i(t,k_j)+\frac{4\pi}{c}j^i(t,k_j),\\
    \i k_i D^i(t, k_j)=4 \pi\rho(t,k_j),\\
    \i k_i B^i(t, k_j)=0.
  \end{gathered}
\end{equation}

Because of the complex form of the resulting equations the complex form of
Maxwell's equations~\eqref{eq:complex:eqiv} is recommended to use
\begin{equation}
  \label{eq:m:fourie:complex}
  \begin{gathered}
    F^i(t,x^j)=\frac{1}{\sqrt{(2\pi)^3}}\int \d^3 k_j \sqrt{\Hat{g}^{(3)}}
    \Hat{F}^i(t,k_j) \e^{\i k_jx^j},\\
    G^i(t,x^j)=\frac{1}{\sqrt{(2\pi)^3}}\int \d^3 k_j \sqrt{\Hat{g}^{(3)}}
    \Hat{G}^i(t,k_j) \e^{\i k_jx^j},\\
    \rho(t,x^j)=\frac{1}{\sqrt{(2\pi)}}\int \d^3 k_j \sqrt{\Hat{g}^{(3)}}
    \Hat{\rho}(t,k_j) \e^{\i k_jx^j}, \\
    j^i(t,x^j)=\frac{1}{\sqrt{(2\pi)^3}}\int \d^3 k_j \sqrt{\Hat{g}^{(3)}}
    \Hat{\jmath}^i(t,k_j) \e^{\i k_jx^j}.
  \end{gathered}
\end{equation}

\textbf{Remark.} 
  In terms of classical electrodynamics vectors $E^j$, $H^j$, $B^j$,
  $D^j$ decompositions in wavevector $k^j$ Fourier series correspond
  to these  vectors decomposition in momentum Fourier series in quantum
  mechanics.  That is why the representation \eqref{eq:maxwell:k} may
  be considered as momentum representation.

\subsection{Spinor Form of Maxwell's Equations}
\label{sec:spinor}

The tensor of electromagnetic field $F_{\alpha\beta}$ and its
components $F_{\crd{\alpha}\crd{\beta}}$, $\crd{\alpha},
\crd{\beta}=\overline{0,3}$ may be considered in spinor form
\cite{penrose-rindler-1987::en} (and similarly
for \(G_{\alpha\beta}\)):
\begin{equation}
\label{eq:spinor:F_as_spinor}
\begin{gathered}
F_{\alpha\beta} = F_{AA'BB'}; \\
F_{\crd{\alpha}\,\crd{\beta}}=F_{\crd{A}\,\crd{A}' \crd{B}\,\crd{B}'}
\tensor{g}{_{\crd{\alpha}}}{^{\crd{A}\,\crd{A}'}}\tensor{g}{_{\crd{\beta}}}{^{\crd{B}\,\crd{B}'}}, \\
\crd{A}, \crd{A}', \crd{B}, \crd{B}' =\overline{0,1},\quad \crd{\alpha},
\crd{\beta}=\overline{0,3},
\end{gathered}
\end{equation}
where $\tensor{g}{_{\crd{\alpha}}}{^{\crd{A}\,\crd{A}'}}$,
$\crd{\alpha}=\overline{0,3}$,~--- Infeld--Van der Waerden symbols 
defined in real spinor basis $\varepsilon_{\crd{A}\,\crd{B}}$ in
the following way~\cite{penrose-rindler-1987::en}:
\begin{equation}
\label{eq:g-iv}
\tensor{g}{_{\crd{\alpha}}}{^{\crd{A}\,\crd{A}'}}:=\tensor{g}{_{\crd{\alpha}}}{^{\alpha}}
\tensor{\varepsilon}{_{A}}{^{\crd{A}}} \tensor{\varepsilon}{_{A'}}{^{\crd{A}'}},\quad 
\tensor{g}{_{\crd{A}\,\crd{A}'}}{^{\crd{\alpha}}}:=\tensor{g}{^{\crd{\alpha}}}{_{\alpha}}
\tensor{\varepsilon}{^{A}}{_{\crd{A}}} \tensor{\varepsilon}{^{A'}}{_{\crd{A}'}},
\end{equation}
\begin{equation}
\label{spin-basis}
\varepsilon_{\crd{A}\,\crd{B}}=\varepsilon_{\crd{A}'\,\crd{B}'}=\begin{pmatrix}
0 & 1\\
-1 & 0
\end{pmatrix},\quad
\tensor{\varepsilon}{_{\crd{A}}}{^A}
\tensor{\varepsilon}{_A}{^{\crd{B}}}= \tensor{\varepsilon}{_{\crd{A}}}{^{\crd{B}}}=\begin{pmatrix}
1 & 0\\
0 & 1
\end{pmatrix}
.
\end{equation}

Let's write Maxwell's equations using the spinors.

 The tensor \(F_{\alpha\beta}\) is real and antisymmetric, it
can be represented in the form
\begin{gather}
  \label{eq:F_spinor}
  F_{\alpha\beta} = \varphi_{AB} \varepsilon_{A'B'} + \varepsilon_{AB}
  \Bar{\varphi}_{A'B'}, \\
  \label{eq:*F^spinor}
  \prescript{*}{}{F}^{\alpha\beta} = - \i \varphi^{AB}\varepsilon^{A'B'} +
  \i \varepsilon^{AB} \Bar{\varphi}^{A'B'}.
\end{gather}
where $\varphi_{AB}$ is a spinor of electromagnetic field:
\[
\varphi_{AB} \vcentcolon= \frac{1}{2}\tensor{F}{_{ABC'}}{^{C'}}=\frac{1}{2}F_{AA'BB'}\varepsilon^{A'B'}=\frac{1}{2}F_{\alpha\beta}\varepsilon^{A'B'}. 
\]

Similarly
\begin{gather}
  \label{eq:G^spinor}
  G^{\alpha\beta} = \gamma^{AB}\varepsilon^{A'B'} +
  \varepsilon^{AB} \Bar{\gamma}^{A'B'}, \\
  \label{eq:*G_spinor}
  \prescript{*}{}{G}_{\alpha\beta} = - \i \gamma_{AB}\varepsilon_{A'B'} +
  \i \varepsilon_{AB} \Bar{\gamma}_{A'B'}.
\end{gather}

Replacing in \eqref{eq:m:tensor} abstract indices $\alpha$ by $AA'$
and $\beta$ by $BB'$, we can write:
\[
\nabla_{AA'} G^{AA' BB'}=\frac{4 \pi}{c}j^{BB'}.
\]

Using~\eqref{eq:G^spinor} we will get 
\begin{equation}
  \label{eq:spinor:maxwell:G}
  \nabla^{AB'} \gamma^{B}_{A} + \nabla^{BA'} \gamma^{B'}_{A'} =
  \frac{4 \pi}{c}j^{BB'}.
\end{equation}

Similarly, from \eqref{eq:m:tensor:2:dual} and \eqref{eq:*F^spinor}
it follows
\begin{equation}
  \label{eq:spinor:maxwell:*G}
  \nabla^{A'B} \varphi^{A}_{B} - \nabla^{AB'} \Bar{\varphi}^{A'}_{B'} = 0.
\end{equation}

  In so doing  the system of Maxwell’s equations can be written as 
\begin{equation}
  \label{eq:spinor:maxwell}
  \begin{gathered}
    \nabla^{A'B} \varphi^{A}_{B} - \nabla^{AB'} \Bar{\varphi}^{A'}_{B'}
    = 0, \\
    \nabla^{AB'} \gamma^{B}_{A} + \nabla^{BA'} \gamma^{B'}_{A'} =
    \frac{4 \pi}{c}j^{BB'}.
  \end{gathered}
\end{equation}

The spinor form of Maxwell's equations system in vacuum can be written
in the form of one equation
\begin{equation}
  \label{eq:maxwell-spin}
  \nabla^{AB'} \varphi^B_A = \frac{2 \pi}{c}j^{BB'}.
  %
\end{equation}

The components of electromagnetic field spinor:
\begin{equation}
\begin{gathered}
\varphi_{\crd{A}\,\crd{B}}=\frac{1}{2}F_{\crd{\alpha}\crd{\beta}}\varepsilon^{\crd{A'}\,\crd{B'}} 
\tensor{g}{^{\crd{\alpha}}}{_{\crd{A}\,\crd{A'}}}
\tensor{g}{^{\crd{\beta}}}{_{\crd{B}\,\crd{B'}}},  \\
\crd{A}, \crd{A}', \crd{B}, \crd{B}' =\overline{0,1},\quad \crd{\alpha},
\crd{\beta}=\overline{0,3}.
\end{gathered}
\end{equation}

Using the equations \eqref{eq:g-iv}, \eqref{spin-basis} and notation
$F_{i}=E_{i} - \i B^{i}$, we will
get~\cite{penrose-rindler-1987::en}:
\begin{equation}
\begin{gathered}
\varphi_{00}=\frac{1}{2}\left(F_1 - \i F_2\right), \\
\varphi_{01}=\varphi_{10}=-\frac{1}{2} F_3,\\
\varphi_{11}=-\frac{1}{2}\left( F_1 + \i F_2\right).
\end{gathered}
\end{equation}

  \section{Maxwell's Equations Presentation in Some Coordinate Systems}
\label{sec:realisation}

\subsection{Maxwell's Equations in Cylindric Coordinate System}

Due to the standard ISO~31-11 the coordinates $(x^{1}, x^{2}, x^{3})$ are
denoted as $(\rho, \varphi, z)$.  In order to avoid some collisions
with charge density symbol $\rho$ the following notation $(r, \varphi,
z)$ will be used.

The law of coordinate transition from Cartesian coordinates to cylindric ones:
\begin{equation}
  \begin{cases}
    x=r\cos\varphi, \\
    y=r\sin\varphi, \\
    z=z.
  \end{cases}
\end{equation}

The law of coordinate transition from cylindric coordinates to Cartesian ones:
\begin{equation}
  \begin{cases} 
    r=\sqrt{x^2+y^2}, \\
    \varphi=\arctg \left(\dfrac{y}{x}\right), \\
    z=z.
  \end{cases}
\end{equation}

The metric tensor:
\begin{equation}
  g_{\crd{i}\crd{j}} =
  \begin{pmatrix} 
    1 & 0 & 0\\ 
    0 & r^2 & 0\\ 
    0 & 0 & 1 
  \end{pmatrix},\quad 
  g^{\crd{i}\crd{j}} = 
  \begin{pmatrix} 
    1 & 0 & 0\\ 
    0 & 1/r^2 & 0\\ 
    0 & 0 & 1 
  \end{pmatrix}.
\end{equation}

\begin{equation}
  \sqrt{g} = r.
\end{equation}

Lame coefficients:
\begin{equation}
  h_{1} \equiv h_r=1, \quad h_{2} \equiv h_\varphi=r, \quad h_{3} \equiv h_z=1.
\end{equation}

The relation between the holonomic (tensor) and nonholonomic (vector)
bases (see~\eqref{eq:vec_nonholonom} and \eqref{eq:covec_nonholonom}):
\begin{gather}
f^{{r}'} = f^{{r}}, \quad
f^{{\varphi}'} = r f^{{\varphi}}, \quad
f^{{z}'} = f^{{z}}, \\
f_{{r}'} = f_{{r}}, \quad
f_{{\varphi}'} = \frac{1}{r} f_{{\varphi}}, \quad
f_{{z}'} = f_{{z}}.
\end{gather}

Differential operators in the holonomic basis:

\begin{gather}
  (\Grad f)_i= 
  \frac{\partial f}{\partial r} \delta_i^{{r}}
  + \frac{\partial f}{\partial \varphi} \delta_i^{{\varphi}} 
  + \frac{\partial f}{\partial z} \delta_i^{{z}}; \\
  \Div \vec{f} = 
  \frac{1}{r} \partial_{r}\left(r f^{{r}}\right)
  +  \partial_{\varphi}\left(f^{{\varphi}}\right)
  + \partial_{z}\left(f^{{z}}\right); \\
  \begin{multlined}
    \left(\Rot \vec{f} \right)^{i} = \frac{1}{r}
    \left[ \partial_{\varphi} f_{{z}} - \partial_{z}
      f_{{\varphi}} \right] \delta_{{r}}^{i} + {} \\
    {} +
    \frac{1}{r}
    \left[ \partial_{z} f_{{r}} - \partial_{r} f_{{z}} \right]
    \delta_{{\varphi}}^{i} + \frac{1}{r} \left[ \partial_{r}
      f_{{\varphi}} - \partial_{\varphi} f_{{r}} \right]
    \delta_{{z}}^{i}.
  \end{multlined}
\end{gather}

Differential operators in the nonholonomic basis:

\begin{gather}
  (\Grad f)_i= 
  \frac{\partial f}{\partial r} \delta_i^{{r}'}
  + \frac{1}{r}
  \frac{\partial f}{\partial \varphi} \delta_i^{{\varphi}'} 
  + \frac{\partial f}{\partial z} \delta_i^{{z}'}; \\
  \Div \vec{f} = 
  \frac{1}{r} \partial_{r}\left(r f^{{r}'}\right)
  + \frac{1}{r} \partial_{\varphi}\left(f^{{\varphi}'}\right)
  + \partial_{z}\left(f^{{z}'}\right); \\
  \begin{multlined}
    \left(\Rot \vec{f} \right)^{i} = \frac{1}{r}
    \left[ \partial_{\varphi} f_{{z}'} - r \partial_{z}
     f_{{\varphi}'} \right] \delta_{{r}'}^{i} + {} \\
    {} +
    \left[ \partial_{z} f_{{r}'} - \partial_{r} f_{{z}'} \right]
    \delta_{{\varphi}}^{i} + 
    \frac{1}{r} \left[ \partial_{r}
      (r f_{{\varphi}'}) - \partial_{\varphi} f_{{r}'} \right]
    \delta_{{z}'}^{i}.
  \end{multlined}
\end{gather}

Maxwell's equations in cylindric coordinates $(r,\varphi,z)$:
\begin{equation}
  \label{eq:maxwell_tc:1}
  \begin{gathered}
    \frac{1}{r}\left[\partial_{\crd{j}}E_{\crd{k}}-\partial_{\crd{k}}E_{\crd{j}}\right]
    = -\frac{1}{c}\partial_t B^{\crd{i}}, \quad \crd{i},\,\crd{j},\,
    \crd{k}=\overline{1,3},\\
    \frac{1}{r}\left[\partial_{\crd{j}}H_{\crd{k}}-\partial_{\crd{k}}H_{\crd{j}}\right]
    = -\frac{1}{c}\partial_t D^{\crd{i}}+\frac{4\pi}{c}j^{\crd{i}}, \quad \crd{i},\,\crd{j},\,
    \crd{k}=\overline{1,3},\\
\frac{1}{r}\partial_{\crd{i}}\left(r
  D^{\crd{i}}\right)= 4\pi \rho, \quad \crd{i}=\overline{1,3},\\
\frac{1}{r}\partial_{\crd{i}}\left(r
  B^{\crd{i}}\right)= 0, \quad \crd{i}=\overline{1,3}.
  \end{gathered}
\end{equation}

The final result after some rearrangements:
\begin{gather}
  \label{eq:maxwell_tc:2}
    \frac{1}{r}\left[\partial_{\varphi}E_3
      - \partial_zE_2\right]=-\frac{1}{c}\partial_tB^1,\\
    \frac{1}{r}\left[\partial_{z}E_1
      - \partial_rE_3\right]=-\frac{1}{c}\partial_tB^2,\\
    \frac{1}{r}\left[\partial_{r}E_2
      - \partial_\varphi E_1\right]=-\frac{1}{c}\partial_tB^3,\\
    \frac{1}{r}\left[\partial_{\varphi}H_3
      - \partial_zH_2\right]=-\frac{1}{c}\partial_tD^1+ \frac{4\pi}{c}j^1,\\
    \frac{1}{r}\left[\partial_{z}H_1
      - \partial_rH_3\right]=-\frac{1}{c}\partial_tD^2+ \frac{4\pi}{c}j^2,\\
    \frac{1}{r}\left[\partial_{r}H_2
      - \partial_\varphi H_1\right]=-\frac{1}{c}\partial_tD^3+ \frac{4\pi}{c}j^3,\\
    \frac{1}{r}D^1+\frac{\partial D^1}{\partial r}+\frac{\partial
      D^2}{\partial \varphi}+\frac{\partial D^3}{\partial z}=4 \pi\rho,\\
    \frac{1}{r}B^1+\frac{\partial B^1}{\partial r}+\frac{\partial
      B^2}{\partial \varphi}+\frac{\partial B^3}{\partial z}=0.
\end{gather}

\subsection{Maxwell's equations in Spherical Coordinate System}

Due to standard ISO~31-11 coordinates $(x^{1}, x^{2}, x^{3})$ are
denoted as $(r, \vartheta, \varphi)$.

The law of coordinate transition from Cartesian coordinates to
spherical ones:
\begin{equation}
  \begin{cases} 
    x=r\sin\vartheta\cos\varphi, \\
    y=r\sin\vartheta\sin\varphi, \\ 
    z=r\cos\vartheta. \end{cases}
\end{equation}
  
The law of coordinate transition from spherical coordinates to Cartesian ones:
\begin{equation}
  \begin{cases} 
    r=\sqrt{x^2+y^2+z^2}, \\
    \vartheta=\arccos \left({\dfrac{z}{\sqrt{x^2+y^2+z^2}}}\right) =
    \arctg \left({\dfrac{\sqrt{x^2+y^2}}{z}}\right), \\ 
    \varphi=\arctg\left({\dfrac{y}{x}}\right). 
  \end{cases} 
\end{equation}

The metric tensor:
\begin{equation}
  g_{\crd{i}\crd{j}} = 
  \begin{pmatrix} 
    1 & 0 & 0\\ 
    0 & r^2 & 0\\ 
    0 & 0 & r^2\sin^2\vartheta 
  \end{pmatrix}, \quad 
  g^{\crd{i}\crd{j}} =
  \begin{pmatrix} 
    1 & 0 & 0\\ 0 & 
    \dfrac{1}{r^2} & 0\\ 0 & 0 &
    \dfrac{1}{r^2\sin^2\vartheta} 
  \end{pmatrix} 
\end{equation}
  
\begin{equation}
  \sqrt{g} = r^2 \sin \vartheta.
\end{equation}

Lame coefficients:
\begin{equation}
  h_{1} \equiv h_r=1, \quad h_{2} \equiv h_\vartheta=r, \quad h_{3} \equiv h_\varphi=r\sin\vartheta.
\end{equation}

The relation between the holonomic (tensor) and nonholonomic (vector)
bases (see~\eqref{eq:vec_nonholonom} and \eqref{eq:covec_nonholonom}):
\begin{gather}
f^{{r}'} = f^{{r}}, \quad
f^{{\vartheta}'} = r f^{{\vartheta}}, \quad
f^{{\varphi}'} = r\sin\vartheta f^{{\varphi}}, \\
f_{{r}'} = f_{{r}}, \quad
f_{{\vartheta}'} = \frac{1}{r} f_{{\vartheta}}, \quad
f_{{\varphi}'} = \frac{1}{r\sin\vartheta} f_{{\varphi}}.
\end{gather}

Differential operators in the holonomic basis:

\begin{gather}
  (\Grad f)_i= 
  \frac{\partial f}{\partial r} \delta_i^{{r}}
  + \frac{\partial f}{\partial \vartheta} \delta_i^{{\vartheta}} 
  + \frac{\partial f}{\partial \varphi} \delta_i^{{\varphi}}; \\
  \Div \vec{f} = 
  \frac{1}{r^2} \partial_{r}\left(r^2 f^{{r}}\right)
  +  \frac{1}{\sin\vartheta}
  \partial_{\vartheta}\left(\sin\vartheta f^{{\vartheta}}\right)
  + \partial_{\varphi}\left(f^{{\varphi}}\right); \\
  \begin{multlined}
    \left(\Rot \vec{f} \right)^{i} = \frac{1}{r^2 \sin\vartheta}
    \left[ \partial_{\vartheta} f_{{\varphi}} - \partial_{\varphi}
      f_{{\vartheta}} \right] \delta_{{r}}^{i} + {} \\
    {} +
    \frac{1}{r^2 \sin\vartheta}
    \left[ \partial_{\varphi} f_{{r}} - \partial_{r} f_{{\varphi}} \right]
    \delta_{{\vartheta}}^{i} + \frac{1}{r^2 \sin\vartheta} \left[ \partial_{r}
      f_{{\vartheta}} - \partial_{\vartheta} f_{{r}} \right]
    \delta_{{\varphi}}^{i}.
  \end{multlined}
\end{gather}

Differential operators in the nonholonomic basis:

\begin{gather}
  (\Grad f)_i= 
  \frac{\partial f}{\partial r} \delta_i^{{r}'}
  + \frac{1}{r}
  \frac{\partial f}{\partial \vartheta} \delta_i^{{\vartheta}'} 
  + \frac{1}{r\sin\vartheta}
  \frac{\partial f}{\partial \varphi} \delta_i^{{\varphi}'}; \\
  \begin{multlined}
    \Div \vec{f} = 
    \frac{1}{r^2} \partial_{r}\left(r^2 f^{{r}'}\right)
    +  \frac{1}{r\sin\vartheta}
    \partial_{\vartheta}\left(\sin\vartheta
      f^{{\vartheta}'}\right) + {} \\ {}
    + \frac{1}{r\sin\vartheta}
    \partial_{\varphi}\left(f^{{\varphi}'}\right); 
  \end{multlined}
  \\
  \begin{multlined}
    \left(\Rot \vec{f} \right)^{i} = \frac{1}{r \sin\vartheta}
    \left[ \partial_{\vartheta}(\sin\vartheta f_{{\varphi}'}) - \partial_{\varphi}
      f_{{\vartheta}'} \right] \delta_{{r}'}^{i} + {} \\
    {} +
    \frac{1}{r}
    \left[ \frac{1}{\sin\vartheta} \partial_{\varphi} f_{{r}'}
      - \partial_{r} (r f_{{\varphi}'}) \right]
    \delta_{{\vartheta}'}^{i} + \frac{1}{r} \left[ \partial_{r}
     (r f_{{\vartheta}'}) - \partial_{\vartheta} f_{{r}'} \right]
    \delta_{{\varphi}'}^{i}.
  \end{multlined}
\end{gather}

Maxwell's equations in spherical coordinates $(r,\vartheta,\varphi)$:
\begin{equation}
  \label{eq:maxwell_sf:1}
  \begin{gathered}
    \frac{1}{r^2\sin\vartheta}\left[\partial_{\crd{j}}E_{\crd{k}}-\partial_{\crd{k}}E_{\crd{j}}\right]
    = -\frac{1}{c}\partial_t B^{\crd{i}}, \quad \crd{i},\,\crd{j},\,
    \crd{k}=\overline{1,3},\\
    \frac{1}{r^2\sin\vartheta}\left[\partial_{\crd{j}}H_{\crd{k}}-\partial_{\crd{k}}H_{\crd{j}}\right]
    = -\frac{1}{c}\partial_t D^{\crd{i}}+\frac{4\pi}{c}j^{\crd{i}}, \quad \crd{i},\,\crd{j},\,
    \crd{k}=\overline{1,3},\\
\frac{1}{r^2\sin\vartheta}\partial_{\crd{i}}\left(r^2\sin\vartheta
  D^{\crd{i}}\right)= 4\pi \rho, \quad \crd{i}=\overline{1,3},\\
\frac{1}{r^2\sin\vartheta}\partial_{\crd{i}}\left(r^2\sin\vartheta
  B^{\crd{i}}\right)= 0, \quad \crd{i}=\overline{1,3}.
  \end{gathered}
\end{equation}

The final result after some rearrangements:
\begin{gather}
  \label{eq:maxwell_sf:2}
    \frac{1}{r^2\sin\vartheta}\left[\partial_{\vartheta}E_3
      - \partial_\varphi E_2\right]=-\frac{1}{c}\partial_tB^1,\\
    \frac{1}{r^2\sin\vartheta}\left[\partial_{\varphi}E_1
      - \partial_rE_3\right]=-\frac{1}{c}\partial_tB^2,\\
    \frac{1}{r^2\sin\vartheta}\left[\partial_{r}E_2
      - \partial_\vartheta E_1\right]=-\frac{1}{c}\partial_tB^3,\\
    \frac{1}{r^2\sin\vartheta}\left[\partial_{\vartheta}H_3
      - \partial_\varphi H_2\right]=-\frac{1}{c}\partial_tD^1+ \frac{4\pi}{c}j^1,\\
    \frac{1}{r^2\sin\vartheta}\left[\partial_{\varphi}H_1
      - \partial_rH_3\right]=-\frac{1}{c}\partial_tD^2+ \frac{4\pi}{c}j^2,\\
    \frac{1}{r^2\sin\vartheta}\left[\partial_{r}H_2
      - \partial_\vartheta H_1\right]=-\frac{1}{c}\partial_tD^3+ \frac{4\pi}{c}j^3,\\
    \frac{2}{r}D^1+\partial_rD^1+\ctg \vartheta D^2+\partial_\vartheta
    D^2+\partial_\varphi D^3=4\pi\rho,\\
    \frac{2}{r}B^1+\partial_rB^1+\ctg \vartheta B^2+\partial_\vartheta
    B^2+\partial_\varphi B^3=0.
\end{gather}

\section{Conclusion}

The main results of the article are:
\begin{enumerate}
\item The connection between tensorial and vectorial formalisms is
  shown.
\item The covariant coordinate representation of differential
  operators for holonomic coordinate systems is given.
\item It is shown how to use tensor formalism for different forms of
  Maxwell's equations.
\item Maxwell's equations are presented in covariant coordinate-free
  and covariant coordinate forms.
\item It is shown that the results obtained by tensorial and vectorial
  formalisms are the same for cylindrical and spherical coordinate
  systems.
\item It is shown that the usage of tensorial formalism instead of
  vectorial one for Maxwell's equations may simplify mathematical
  expressions (particularly in non-Cartesian coordinate systems).

  Using tensorial formalism instead of vectorial one can simplify the
  form of equations and intermediate results in non-Cartesian
  coordinate systems due to well developed formalism of tensor
  analysis.  The transition to vectorial formalism can be done at a
  final stage if necessary.
\end{enumerate}

\bibliographystyle{abbrvnat}

\bibliography{bib/ref}

\begin{thebibliography}{6}
\providecommand{\natexlab}[1]{#1}
\providecommand{\url}[1]{\texttt{#1}}
\expandafter\ifx\csname urlstyle\endcsname\relax
  \providecommand{\doi}[1]{doi: #1}\else
  \providecommand{\doi}{doi: \begingroup \urlstyle{rm}\Url}\fi

\bibitem[Kulyabov and Nemchaninova(2011)]{PFUR-2011-2-kul::en}
D.~S. Kulyabov and N.~A. Nemchaninova.
\newblock Maxwell’s equations in curvilinear coordinates.
\newblock \emph{Bulletin of Peoples’ Friendship University of Russia. Series
  Mathematics. Information Sciences. Physics}, \penalty0 (2):\penalty0
  172--179, 2011.

\bibitem[Stratton(1941)]{stratton:1948::en}
J.~A. Stratton.
\newblock \emph{Electromagnetic Theory}.
\newblock MGH, 1941.

\bibitem[Penrose and Rindler(1984)]{penrose-rindler-1987::en}
R.~Penrose and W.~Rindler.
\newblock \emph{Spinors and Space-Time: Two-Spinor Calculus and Relativistic
  Fields}, volume~1.
\newblock Cambridge University Press, 1984.

\bibitem[Minkowski(1910)]{minkowski:1910::en}
H.~Minkowski.
\newblock Die grundlagen f\"{u}r die electromagnetischen vorg\"{o}nge in
  bewegten k\"{o}rpern.
\newblock \emph{Math. Ann.}, \penalty0 (68):\penalty0 472--525, 1910.

\bibitem[Silberstein(1907)]{silberstein:1907::en}
L.~Silberstein.
\newblock Electromagnetische grundgleichungen in bivectorieller behandlung.
\newblock \emph{Annalen der Physik}, 22:\penalty0 579--586, 1907.

\bibitem[Morse and Feshbach(1953)]{mors::en}
F.~M. Morse and H.~Feshbach.
\newblock \emph{Methods of theoretical physics}, volume~1.
\newblock MGH, 1953.

\end{thebibliography}


\begin{thebibliography}{1}
\def\selectlanguageifdefined#1{
\expandafter\ifx\csname date#1\endcsname\relax
\else\selectlanguage{#1}\fi}
\providecommand*{\href}[2]{{\small #2}}
\providecommand*{\url}[1]{{\small #1}}
\providecommand*{\BibUrl}[1]{\url{#1}}
\providecommand{\BibAnnote}[1]{}
\providecommand*{\BibEmph}[1]{#1}
\ProvideTextCommandDefault{\cyrdash}{\hbox to.8em{--\hss--}}
\providecommand*{\BibDash}{\ifdim\lastskip>0pt\unskip\nobreak\hskip.2em\fi
\cyrdash\hskip.2em\ignorespaces}

\bibitem{PFUR-2011-2-kul}
\selectlanguageifdefined{russian}
\BibEmph{Кулябов~Д.~С., Немчанинова~Н.~А.}
  Уравнения Максвелла в криволинейных
  координатах~// \BibEmph{Вестник РУДН. Серия
  <<Математика. Информатика. Физика>>}. \BibDash
\newblock 2011. \BibDash
\newblock {№}~2. \BibDash
\newblock {С.}~172--179.

\bibitem{penrose-rindler-1987}
\selectlanguageifdefined{russian}
\BibEmph{Пенроуз~Р., Риндлер~В.} Спиноры и
  пространство-время. Два-спинорное
  исчисление и релятивистские поля. \BibDash
\newblock М.~: Мир, 1987. \BibDash
\newblock Т.~1. \BibDash
\newblock 528~{с.}

\bibitem{mors}
\selectlanguageifdefined{russian}
\BibEmph{Морс~Ф.~М., Фешбах~Г.} Методы
  теоретической физики. \BibDash
\newblock М.~: Издательство иностранной
  литературы, 1960.

\bibitem{vasiliev}
\selectlanguageifdefined{russian}
\BibEmph{Васильев~А.~Н.} Классическая
  электродинамика. Краткий курс лекций.
  \BibDash
\newblock C.-П.~: БХВ-Петербург, 2010.

\bibitem{minkowski:1910}
\selectlanguageifdefined{german}
\BibEmph{Minkowski~H.} Die Grundlagen f\"{u}r die electromagnetischen
  Vorg\"{o}nge in bewegten K\"{o}rpern~// \BibEmph{Math. Ann.} \BibDash
\newblock 1910. \BibDash
\newblock {H.}~68. \BibDash
\newblock S.~472--525.

\bibitem{terletskiy-rybakov-1990}
\selectlanguageifdefined{russian}
\BibEmph{Терлецкий~Я.~П., Рыбаков~Ю.~П.}
  Электродинамика: Учебное пособие для
  студентов физ. спец. университетов. \BibDash
\newblock 2-е, перераб. {изд.} \BibDash
\newblock М.~: Высш. шк., 1990. \BibDash
\newblock 352~{с.}

\bibitem{stratton:1948}
\selectlanguageifdefined{russian}
\BibEmph{Стрэттон~{Дж}.~А.} Теория
  электромагнетизма. \BibDash
\newblock М.-Л.: ГИТТЛ, 1948.

\bibitem{silberstein:1907}
\selectlanguageifdefined{german}
\BibEmph{Silberstein~L.} Electromagnetische Grundgleichungen in bivectorieller
  Behandlung~// \BibEmph{Annalen der Physik}. \BibDash
\newblock 1907. \BibDash
\newblock {Bd.}~22. \BibDash
\newblock S.~579--586.

\bibitem{batygin:2002}
\selectlanguageifdefined{russian}
\BibEmph{В.~Батыгин~В., Н.~Топтыгин~И.} Сборник
  задач по электродинамике. \BibDash
\newblock М.~: НИЦ «Регулярная и хаотическая
  динамика», 2002.

\end{thebibliography}

\end{document}


\author{Д. С. Кулябов}
\email{yamadharma@gmail.com}
\author{А. В. Королькова}
\email{avkorolkova@gmail.com}

\affiliation{Российский университет дружбы народов}

\thanks{Опубликовано в: Kulyabov D.S., Korolkova A.V., Korolkov
  V.I. Maxwell’s Equations in Arbitrary Coordinate System // Bulletin
  of Peoples’ Friendship University of Russia. Series
  «Mathematics. Information Sciences. Physics». 2012. №
  1. P.~96--106.}

\title{Уравнения Максвелла в произвольной системе координат}

\begin{abstract}
  В работе продемонстрировано применение тензорного формализма для
  получения разных форм записи уравнений Максвелла. Получены уравнения
  Максвелла в ковариантной бескоординатной и ковариантной координатной
  формах.  Предварительно установлена связь между векторным и
  тензорным формализмами, выписано координатное представление
  дифференциальных операторов для произвольных голономных систем
  координат. Проведена верификация  результатов, полученных с помощью
  тензорного и векторного формализмов на примере цилиндрической и
  сферической систем координат.
\end{abstract}

\maketitle

\section{Введение}

В задачах математического моделирования волноводов часто возникает
потребность использования криволинейных систем координат.  Выбор
конкретной системы координат зависит от формы поперечного сечения
волновода.

Обычно для описания исследуемой волноводной модели за основу берётся
запись уравнений Максвелла в декартовой системе координат. Используя
трансформационные свойства векторов, уравнения Максвелла
переписываются для конкретной криволинейной системы координат, часто
сферической или цилиндрической.  Однако, например, в задаче
моделирования ускорителя тяжёлых частиц волновод может иметь форму
конуса или гиперболоида.  Другой пример волновода сложной формы ---
волноводная линза Люнеберга, представляющая собой часть сферы или
цилиндра, прикреплённой к планарному волноводу.  Поэтому, в
случае более сложной формы волновода требуется запись уравнений
Максвелла в произвольной криволинейной системе координат.

Традиционно к уравнениям Максвелла применяют векторный формализм. В
этом случае запись уравнений в криволинейной системе координат крайне
громоздка.  В работе~\cite{PFUR-2011-2-kul} проведены предварительные
исследования по применению тензорного формализма, приводящего к более
компактной и удобной форме записи уравнений Максвелла. Кроме того,
тензорный формализм имеет мощный математический аппарат, который
позволяет работать с ковариантной бескоординатной формой записи
уравнений. В этом случае переход к конкретной системе координат нужен
только на заключительном этапе исследований при записи результата. 
Непосредственно применить тензорный формализм к уравнениям Максвелла
нельзя. Требуется установить связь векторного и тензорного формализмов.

Различные формы записи уравнений Максвелла используются и в задачах,
связанных с нахождением гамильтониана электромагнитного поля,
применяемого для построения вариационных интеграторов, в частности,
симплектических интеграторов. Основная цель --- выполнение условия
сохранения симплектической структуры при дискретизации уравнений.
Методы нахождения гамильтониана электромагнитного поля используют
следующие формы записи уравнений Максвелла:
\begin{itemize}
\item через 3-векторы;
\item через 4-векторы;
\item комплексное представление;
\item импульсное представление (для его записи в свою очередь
  применяется комплексная форма);
\item спинорное представление.
\end{itemize}

Резюмируя сказанное выше, сформулируем основные задачи, решаемые в
данной работе. Необходимо продемонстрировать связь векторного и
тензорного формализмов (раздел \ref{s.1}).  Применить тензорный
формализм для различных форм представления уравнений Максвелла (раздел
\ref{sec:maxwell_represent}) и верифицировать полученные результаты,
выписав явно уравнения Максвелла в цилиндрической и сферической
системах координат (раздел \ref{sec:realisation}).

\section{Связь тензорного и векторного формализмов записи векторов}
\label{s.1}

  Будем использовать предложенный в~\cite{penrose-rindler-1987}
  формализм абстрактных индексов. В~\cite{penrose-rindler-1987} через
  $\alpha$ обозначен абстрактный индекс, $\crd{\alpha}$~---
  компонентный индекс тензора. Присутствие в некотором выражении
  компонентного индекса означает, что в него косвенным образом введён
  некоторый (произвольный) базис, а сами индексы подчиняются правилу
  суммирования Эйнштейна (суммирование по всякому численному индексу,
  который встречается в одном члене выражения дважды: вверху и
  внизу). Абстрактные индексы имеют организующее значение.

Рассмотрим произвольное 
$n$-мерное
векторное пространство $V^{i}$
и сопряжённое к $V^{i}$ пространство $V_{i}$.

В тензорном формализме зададим голономный базис:
\[
\delta_{\crd{i}}^{i}=\frac{\partial}{\partial x^{\crd{i}}} \in V^{i},\quad 
\delta^{\crd{i}}_{i}=\d x^{\crd{i}} \in V_{i},\quad \crd{i}=\overline{1,n}.
\]

В векторном формализме неголономный базис задаётся через элементы длины $\d
s^{\crd{i}'}$ по соответствующей координате:
\[
\delta_{\crd{i}'}^{i}=\frac{\partial}{\partial s^{\crd{i}'}},\quad 
\delta^{\crd{i}'}_{i}=\d s^{\crd{i}'},\quad \crd{i}'=\overline{1,n}.
\]

Неголономный базис обычно предоставляет некоторые удобства. В данном
случае это:
\begin{itemize}
\item сохранение величин при преобразовании координат
  (т.\,е. расстояния переходят в расстояние, углы в углы и т.\,д.);
\item неразличимость контравариантных и ковариантных векторов, что
  позволяет использовать только один тип индекса.
\end{itemize}

В тензорной записи:
\begin{equation}
\label{eq:ds2-t}
\d s^2= g_{\crd{i}\crd{j}}\d x^{\crd{i}}\d x^{\crd{j}},\quad
\crd{i},\,\crd{j}=\overline{1,n},
\end{equation}

где $g_{\crd{i}\crd{j}}$~--- метрический тензор.

В векторной записи:
\begin{equation}
\label{eq:ds2-v}
\d s^2=g_{\crd{i}' \crd{j}'}\d s^{\crd{i}'}\d s^{\crd{j}'},
\quad \crd{i}',\, \crd{j}'=\overline{1,n}.
\end{equation}

В случае ортогонального базиса \eqref{eq:ds2-v} принимает вид:
\begin{equation}
\label{eq:ds2-v2}
\d s^2=g_{\crd{i}'\crd{i}'}\d s^{\crd{i}'}\d s^{\crd{i}'}, \quad
\crd{i}'=\overline{1,n}. 
\end{equation}

Выразим векторный базис через тензорный:
\[
\d s^{\crd{i}'}=h_{\crd{i}}^{\crd{i}'}\d x^{\crd{i}},\quad 
\frac{\partial}{\partial
  s^{\crd{i}'}}=h_{\crd{i}'}^{\crd{i}}\frac{\partial}{\partial
  x^{\crd{i}}}. 
\]

Здесь $h_{\crd{i}}^{\crd{i}'}$, $h_{\crd{i}'}^{\crd{i}}$,
$\crd{i},\,\crd{i}'=\overline{1,n}$,~--- коэффициенты неголономности.

Для ортогонального базиса из \eqref{eq:ds2-v2} находим:
\[
g_{\crd{i}\crd{i}}\d x^{\crd{i}}\d
x^{\crd{i}}= g_{\crd{i}'\crd{i}'}
h_{\crd{i}}^{\crd{i}'}h_{\crd{i}}^{\crd{i}'}\d x^{\crd{i}}\d
x^{\crd{i}}, \quad \crd{i},\,\crd{i}'=\overline{1,n}.
\]

Введём обозначение (для ортогональной системы координат):
\[
\left(h_{\crd{i}}\right)^{2}:= 
h_{\crd{i}}^{\crd{i}'}h_{\crd{i}}^{\crd{i}'} =
\frac{g_{\crd{i}\crd{i}}}{g_{\crd{i}'\crd{i}'}},\quad 
h_{\crd{i}} := h_{\crd{i}}^{\crd{i}'} =
\sqrt{\frac{g_{\crd{i}\crd{i}}}{g_{\crd{i}'\crd{i}'}}},
\quad
\crd{i},\,\crd{i}'=\overline{1,n}.
\]

Величины $h_{\crd{i}}$ называются коэффициентами Ламе~\cite[Т.~1,
с.~34--35]{mors}.

Выразим вектор $f^{i}\in V^{i}$ через его компоненты $f^{\crd{i}}$ в тензорном
$\delta_{\crd{i}}^{i}$ и векторном $\delta_{\crd{i}'}^{i}$ базисах
соответственно:
\[
\begin{gathered}
f^{i}=f^{\crd{i}}\delta_{\crd{i}}^{i}=f^{\crd{i}}\frac{\partial} 
{\partial x^{\crd{i}}},\\
f^{i}=f^{\crd{i}'}\delta_{\crd{i}'}^{i}=f^{\crd{i}'}\frac{\partial}{\partial
s^{\crd{i}'}}=f^{\crd{i}'}\frac{1}{h_{\crd{i}}^{\crd{i}'}}\frac{\partial}{\partial
x^{\crd{i}}}.
\end{gathered}
\]
Отсюда получаем, что 
\begin{equation}
\label{eq:vec_nonholonom}
f^{\crd{i}'}=f^{\crd{i}} h_{\crd{i}}^{\crd{i}'}, \quad \crd{i},\,\crd{i}'=\overline{1,n}.
\end{equation}

Аналогично для ковекторов имеем:
\[
\begin{gathered}
f_{i}=f_{\crd{i}}\delta^{\crd{i}}_{i}=f_{\crd{i}}\d x^{\crd{i}},\\
f_{i}=f_{\crd{i}'}\delta^{\crd{i}'}_{i}=f_{\crd{i}'}\d
s^{\crd{i}'}=f_{\crd{i}'}h_{\crd{i}}^{\crd{i}'}
\d x^{\crd{i}},
\end{gathered}
\]
откуда получаем, что 
\begin{equation}
\label{eq:covec_nonholonom}
f_{\crd{i}'}=f_{\crd{i}} \frac{1}{h_{\crd{i}}^{\crd{i}'}},\quad \crd{i},\,\crd{i}'=\overline{1,n}.
\end{equation}

Таким образом, показана связь между тензорным и векторным
формализмами.

\section{Тензорная запись дифференциальных операторов в компонентах}
\label{s.2}
Запишем в компонентах дифференциальные операторы (для связностей,
ассоциированных с метрикой).

Выражение для градиента имеет вид (здесь $\varphi$~--- скаляр):
\begin{equation}
\label{eq:grad}
\begin{gathered}
(\Grad \varphi)_i= (\Grad \varphi)_{\crd{i}}
\delta_{i}^{\crd{i}},\\ 
(\Grad \varphi)_{\crd{i}}=\nabla_{\crd{i}}\varphi
=\partial_{\crd{i}}\varphi, \quad \crd{i}=\overline{1,n}.
\end{gathered}
\end{equation}

Выражение для дивергенции некоторого произвольного вектора $\vec{f}\in
V^{i}$ имеет вид:
\begin{multline}
  \label{eq:div}
\Div  \vec{f}= \nabla_i f^i= f_{, i}^i- \Gamma_{ji}^i f^j= f_{, i}^i-
f^i\frac{(\sqrt{|g|})_{, i}}{\sqrt{|g|}}= {} \\
{} = \frac{1}{\sqrt{|g|}}\partial_i\left(\sqrt{|g|} f^i\right),
\end{multline}
или в компонентах:
\begin{equation}
  \label{eq:div_comp}
\Div  \vec{f}=\frac{1}{\sqrt{|g|}}\partial_{\crd{i}}\left(\sqrt{|g|}
  f^{\crd{i}}\right), \quad \crd{i}=\overline{1,n}.
\end{equation}

Здесь $g$ представляет собой $\det\left(g_{\crd{i} \crd{i}}\right)$.
Так как подкоренное выражение должно быть неотрицательным,
а в пространстве Минковского $\mathbb{M}^4$ $g < 0$,
то для определённости будем использовать запись $|g|$.

Выражение для ротора действительно только в пространстве
$\mathbb{E}^3$:
\begin{equation}
\label{eq:rot}
\begin{gathered}
\left(\Rot \vec{f}\right)^{i}=\left[\vec{\nabla}, \vec{f}\right]=
\left(\Rot \vec{f}\right)^{\crd{i}} \delta_{\crd{i}}^{i}, \\ 
\left(\Rot \vec{f}\right)^{\crd{i}}=e^{\crd{i}\crd{j}\crd{k}}
\nabla_{\crd{j}} f_{\crd{k}},\quad \crd{i},\, \crd{j},\,
\crd{k}=\overline{1,3},
\end{gathered}
\end{equation}
где $e^{\crd{i}\crd{j}\crd{k}}$~--- альтернирующий тензор, выражающийся через
символ Леви-Чевиты $\varepsilon^{\crd{i}\crd{j}\crd{k}}$:
\[
e_{\crd{i}\crd{j}\crd{k}}=\sqrt{g^{(3)}}\varepsilon_{\crd{i}\crd{j}\crd{k}},\quad
e^{\crd{i}\crd{j}\crd{k}}=\frac{1}{\sqrt{g^{(3)}}}\varepsilon^{\crd{i}\crd{j}\crd{k}},
\quad \crd{i},\, \crd{j},\, \crd{k}=\overline{1,3}.
\]

В общем случае имеем:
\begin{equation}
\label{eq:e_alter}
\begin{gathered}
  e_{\crd{a_1}\crd{a_2}\dots\crd{a_n}} = 
  \sqrt{|g^{(n)}|} \varepsilon_{\crd{a_1}\crd{a_2}\dots\crd{a_n}},\\
  e^{\crd{a_1}\crd{a_2}\dots\crd{a_n}} =
  \frac{\sqrt{|g^{(n)}|}}{g^{(n)}}
  \varepsilon^{\crd{a_1}\crd{a_2}\dots\crd{a_n}}
  = \frac{\sign g^{(n)}}{\sqrt{|g^{(n)}|}}
  \varepsilon^{\crd{a_1}\crd{a_2}\dots\crd{a_n}}, \\
  \crd{a_1},\, \crd{a_2},\dots , \crd{a_n}=\overline{1,n}.
\end{gathered}
\end{equation}

Из выражений \eqref{eq:div} для дивергенции и 
\eqref{eq:grad} для градиента
можно получить лапласиан:
\begin{multline}
  \label{eq:lapl}
\Delta \varphi= \nabla_i\left(\nabla^i
  \varphi\right)=\nabla_i\left(g^{ij}(\Grad \varphi)_j\right)= {} \\
{} = \nabla_i\left(g^{ij}\partial _j
  \varphi\right)=\frac{1}{\sqrt{|g|}}\partial_i\left(\sqrt{|g|}g^{ij}\partial_j
\varphi\right).
\end{multline}

\section{Представления уравнений Максвелла}
\label{sec:maxwell_represent}

Рассмотрим уравнения Максвелла в системе СГС~\cite{vasiliev}:
\begin{equation}
  \label{eq:maxwell}
  \begin{gathered}
    \vec \nabla \times \vec E = - \frac{1}{c} \frac{\partial
      \vec{B}}{\partial t}; \\
    \vec{\nabla}\cdot\vec{D} = 4\pi\rho; \\
    \vec \nabla \times \vec H = \frac{1}{c} \frac{\partial
      \vec{D}}{\partial t} + \frac{4\pi}{c} \vec{j}; \\
    \vec{\nabla}\cdot\vec{B} = 0.
  \end{gathered}
\end{equation}

Здесь $\vec{E}$ и $\vec{H}$~--- напряжённости электрического и
магнитного полей, $\vec{D}$ и $\vec{B}$~--- электрическая и магнитная
индукция соответственно, $\vec{j}$~--- плотность тока, $\rho$~---
плотность заряда, $c$~--- скорость света.

\subsection{Ковариантная запись уравнений Максвелла через 3-векторы}
\label{sec:3-vector}

Запишем уравнения \eqref{eq:maxwell} в явно ковариантной форме
\begin{equation}
  \label{eq:3-vector:cov}
  \begin{gathered}
   e^{ijk} \nabla_{j} E_{k} = - \nabla_{0} B^{i}; \\
    \nabla_{i}D^{i} = 4\pi\rho; \\
    e^{ijk} \nabla_{j} H_{k} = \nabla_{0} D^i + \frac{4\pi}{c} j^{i}; \\
    \nabla_{i} B^{i} = 0.
  \end{gathered}
\end{equation}

Перепишем \eqref{eq:maxwell} в компонентах тензорного формализма,
используя \eqref{eq:rot} и~\eqref{eq:div_comp}: 
\begin{equation}
  \label{eq:maxwell_comp}
  \begin{gathered}
    \frac{1}{\sqrt{g^{(3)}}}\left[\partial_{\crd{j}}E_{\crd{k}}-\partial_{\crd{k}}E_{\crd{j}}\right]
    = -\frac{1}{c}\partial_t B^{\crd{i}}, \quad \crd{i},\,\crd{j},\,
    \crd{k}=\overline{1,3},\\
    \frac{1}{\sqrt{g^{(3)}}}\partial_{\crd{i}}\left(\sqrt{g^{(3)}}
      D^{\crd{i}}\right)= 4\pi \rho, \quad \crd{i}=\overline{1,3},\\
    \frac{1}{\sqrt{g^{(3)}}}\left[\partial_{\crd{j}}H_{\crd{k}}-\partial_{\crd{k}}H_{\crd{j}}\right]
    = -\frac{1}{c}\partial_t D^{\crd{i}}+\frac{4\pi}{c}j^{\crd{i}}, \quad \crd{i},\,\crd{j},\,
    \crd{k}=\overline{1,3},\\
    \frac{1}{\sqrt{g^{(3)}}}\partial_{\crd{i}}\left(\sqrt{g^{(3)}}
      B^{\crd{i}}\right)= 0, \quad \crd{i}=\overline{1,3}.
  \end{gathered}
\end{equation}

\subsection{Ковариантная запись уравнений Максвелла через 4-векторы}
\label{sec:4-vector}

Запишем \eqref{eq:maxwell} через тензоры
электромагнитного поля 
$F_{\alpha\beta}$ и $G_{{\alpha}{\beta}}$\cite{minkowski:1910}, \cite[с. 256,
263--264]{terletskiy-rybakov-1990}:
\begin{gather}
\nabla_{{\alpha}} F_{{\beta}{\gamma}}+ \nabla_{{\beta}}
F_{{\gamma}{\alpha}}+\nabla_{{\gamma}} F_{{\alpha}{\beta}} = 
F_{[\alpha \beta ; \gamma]} = 0,
\label{eq:m:tensor:2}
\\
\nabla_{{\alpha}} G^{{\alpha}{\beta}}=\frac{4 \pi}{c}j^{{\beta}},
\label{eq:m:tensor}
\end{gather}
где тензоры $F_{\alpha\beta}$, $F^{\alpha\beta}$, $G^{\alpha\beta}$ и $G_{\alpha\beta}$ имеют следующие
компоненты
\begin{gather}
F_{\crd{\alpha}\crd{\beta}}=
\begin{pmatrix} 
0 & {E}_1 & {E}_2 & {E}_3 \\ 
-{E}_1 & 0 & -{B}^3 & {B}^2 \\
-{E}_2 & {B}^3 & 0 & -{B}^1 \\ 
-{E}_3 & -{B}^2 & {B}^1 & 0 
\end{pmatrix},
\\
F^{\crd{\alpha}\crd{\beta}}=
\begin{pmatrix} 
0 & -{E}^1 & -{E}^2 & -{E}^3 \\ 
{E}^1 & 0 & -{B}_3 & {B}_2 \\
{E}^2 & {B}_3 & 0 & -{B}_1 \\ 
{E}^3 & -{B}_2 & {B}_1 & 0 
\end{pmatrix},
\\
G^{\crd{\alpha}\crd{\beta}}=
\begin{pmatrix} 
0 & -{D}^1 & -{D}^2 & -{D}^3 \\ 
{D}^1 & 0 & -{H}_3 & {H}_2 \\
{D}^2 & {H}_3 & 0 & -{H}_1 \\ 
{D}^3 & -{H}_2 & {H}_1 & 0 
\end{pmatrix},
\\
G_{\crd{\alpha}\crd{\beta}}=
\begin{pmatrix} 
0 & {D}_1 & {D}_2 & {D}_3 \\ 
-{D}_1 & 0 & -{H}^3 & {H}^2 \\
-{D}_2 & {H}^3 & 0 & -{H}^1 \\ 
-{D}_3 & -{H}^2 & {H}^1 & 0 
\end{pmatrix},
\end{gather}
  $E^{\crd{i}}$, $H^{\crd{i}}$, $\crd{i}=\overline{1,3}$,~---
  компоненты векторов напряжённости электрического и магнитного полей
  соответственно; $D_{\crd{i}}$, $B_{\crd{i}}$,
  $\crd{i}=\overline{1,3}$,~--- компоненты векторов электрической и
  магнитной индукции соответственно\footnote{Следует заметить, что
    именно $B^{i}$ имеет физический смысл напряжённости магнитного
    поля.}.

Уравнение \eqref{eq:m:tensor:2} можно записать в более простом виде
\begin{equation}
  \label{eq:m:tensor:2:dual}
  \nabla_{\alpha} \prescript{*}{}{F}^{\alpha\beta} = 0.
\end{equation}
Здесь введён тензор $\prescript{*}{}{F}^{\alpha\beta}$, дуально сопряжённый
тензору ${F}^{\alpha\beta}$
\begin{equation}
  \label{eq:tilde_g}
  \prescript{*}{}{F}^{\alpha\beta} = \frac{1}{2} e^{\alpha\beta\gamma\delta} F_{\gamma\delta},
\end{equation}
где $e^{\alpha\beta\gamma\delta}$~--- альтернирующий тензор (см.~\eqref{eq:e_alter}):
\begin{equation}
  \label{eq:e4}
  e_{\crd{\alpha}\crd{\beta}\crd{\gamma}\crd{\delta}} = \sqrt{-g}\varepsilon_{\crd{\alpha}\crd{\beta}\crd{\gamma}\crd{\delta}},\quad
  e^{\crd{\alpha}\crd{\beta}\crd{\gamma}\crd{\delta}} = -\frac{1}{\sqrt{-g}}\varepsilon^{\crd{\alpha}\crd{\beta}\crd{\gamma}\crd{\delta}}.
\end{equation}

Аналогично запишем
\begin{equation}
  \label{eq:tilde_other}
  \begin{gathered}
    \prescript{*}{}{F}_{\alpha\beta} = \frac{1}{2}
    e_{\alpha\beta\gamma\delta} F^{\gamma\delta}, \\
    \prescript{*}{}{G}_{\alpha\beta} = \frac{1}{2}
    e_{\alpha\beta\gamma\delta} G^{\gamma\delta}, \\
    \prescript{*}{}{G}^{\alpha\beta} = \frac{1}{2}
    e^{\alpha\beta\gamma\delta} G_{\gamma\delta}.
  \end{gathered}
\end{equation}

Запишем в компонентах:
\begin{equation}
\begin{gathered}
\prescript{*}{}{F}_{\crd{\alpha}\crd{\beta}}=
\sqrt{-g}
\begin{pmatrix} 
0 & {B}_1 & {B}_2 & {B}_3 \\ 
-{B}_1 & 0 & {E}^3 & -{E}^2 \\
-{B}_2 & -{E}^3 & 0 & {E}^1 \\ 
-{B}_3 & {E}^2 & -{E}^1 & 0 
\end{pmatrix},
\\
\prescript{*}{}{F}^{\crd{\alpha}\crd{\beta}}=
\frac{1}{\sqrt{-g}}
\begin{pmatrix} 
0 & -{B}^1 & -{B}^2 & -{B}^3 \\ 
{B}^1 & 0 & {E}_3 & -{E}_2 \\
{B}^2 & -{E}_3 & 0 & {E}_1 \\ 
{B}^3 & {E}_2 & -{E}_1 & 0 
\end{pmatrix},
\\
\prescript{*}{}{G}^{\crd{\alpha}\crd{\beta}}=
\frac{1}{\sqrt{-g}}
\begin{pmatrix} 
0 & -{H}^1 & -{H}^2 & -{H}^3 \\ 
{H}^1 & 0 & {D}_3 & -{D}_2 \\
{H}^2 & -{D}_3 & 0 & {D}_1 \\ 
{H}^3 & {D}_2 & -{D}_1 & 0 
\end{pmatrix},
\\
\prescript{*}{}{G}_{\crd{\alpha}\crd{\beta}}=
\sqrt{-g}
\begin{pmatrix} 
0 & {H}_1 & {H}_2 & {H}_3 \\ 
-{H}_1 & 0 & {D}^3 & -{D}^2 \\
-{H}_2 & -{D}^3 & 0 & {D}^1 \\ 
-{H}_3 & {D}^2 & -{D}^1 & 0 
\end{pmatrix},
\end{gathered}
\end{equation}

Кодифицируем запись используемых тензоров. Для этого 
поставим в соответствие $F_{\crd{\alpha}\crd{\beta}}$ упорядоченную пару
$(E_{\crd{i}}, B^{\crd{i}})$ ($F_{\crd{\alpha}\crd{\beta}} \sim (E_{\crd{i}}, B^{\crd{i}})$) следующим
образом
\begin{equation}
  \label{eq:F_equiv_para}
  F_{0\crd{i}} = E_{i}, \quad F_{\crd{i}\crd{j}} = - \varepsilon_{\crd{i}\crd{j}\crd{k}} B^{\crd{k}}.
\end{equation}

Таким образом можно выписать следующие соответствия
\begin{equation}
  \label{eq:all_equiv}
  \begin{gathered}
    F_{\crd{\alpha}\crd{\beta}} \sim (E_{\crd{i}}, B^{\crd{i}}),
    \qquad 
    F^{\crd{\alpha}\crd{\beta}} \sim (-E^{\crd{i}}, B_{\crd{i}}), \\
    G_{\crd{\alpha}\crd{\beta}} \sim (D_{\crd{i}}, H^{\crd{i}}), 
    \qquad 
    G^{\crd{\alpha}\crd{\beta}} \sim  (-D^{\crd{i}}, H_{\crd{i}}), \\
    \prescript{*}{}{F}_{\crd{\alpha}\crd{\beta}} \sim 
    \sqrt{-g} (B_{\crd{i}}, -E^{\crd{i}}), 
    \qquad 
    \prescript{*}{}{F}^{\crd{\alpha}\crd{\beta}} \sim 
    \frac{1}{\sqrt{-g}} (-B^{\crd{i}}, -E_{\crd{i}}), \\
    \prescript{*}{}{G}_{\crd{\alpha}\crd{\beta}} \sim 
    \sqrt{-g} (H_{\crd{i}}, -D^{\crd{i}}), 
    \qquad 
    \prescript{*}{}{G}^{\crd{\alpha}\crd{\beta}} \sim 
    \frac{1}{\sqrt{-g}} (-H^{\crd{i}}, -D_{\crd{i}}).
  \end{gathered}
\end{equation}

По этой упорядоченной паре можно строить как тензорные, так и векторные
представления тензора электромагнитной индукции.

\subsection{Комплексное представление уравнений Максвелла}
\label{sec:complex}

Можно построить несколько видов комплексного представления уравнений
Максвелла, хотя обычно ограничиваются вакуумным случаем в евклидовом
пространстве. Мы же запишем комплексное представление в среде в
произвольных координатах.

Комплексное представление уравнений Максвелла рассматривалось разными
авторами \cite[с.~40--42]{stratton:1948}, \cite{silberstein:1907}.

Аналогично \eqref{eq:all_equiv} зададим соответствие упорядоченной
пары и комплексного 3-вектора
\begin{equation}
  \label{eq:complex:eqiv}
  \begin{gathered}
    F^{\crd{i}} \sim (E^{\crd{i}}, B^{\crd{i}}), 
    \quad 
    F^{\crd{i}} = E^{\crd{i}} + \i B^{\crd{i}}; \\
    G^{\crd{i}} \sim (D^{\crd{i}}, H^{\crd{i}}), 
    \quad 
    G^{\crd{i}} = D^{\crd{i}} + \i H^{\crd{i}}.
  \end{gathered}
\end{equation}

Выразим напряжённость и индукцию через соответствующие комплексные
векторы
\begin{equation}
  \label{eq:complex:vectors}
  \begin{gathered}
    E^{i} = \frac{F^{i} + \Bar{F}^{i}}{2}, \quad 
    B^{i} = \frac{F^{i} - \Bar{F}^{i}}{2 \i}, \\
    D^{i} = \frac{G^{i} + \Bar{G}^{i}}{2}, \quad 
    H^{i} = \frac{G^{i} - \Bar{G}^{i}}{2 \i}.
  \end{gathered}
\end{equation}

Введём два дополнительных комплексных вектора
\begin{equation}
  \label{eq:complex:K_L}
  K^{i} = \frac{G^{i} + F^{i}}{2}, \quad L^{i} = \frac{\Bar{G}^{i} - \Bar{F}^{i}}{2}.
\end{equation}

Тогда уравнения \eqref{eq:3-vector:cov} примут вид
\begin{equation}
  \label{eq:complex:maxwell_K_L}
  \begin{gathered}
    \nabla_{i} (K^{i} + L^{i}) = 4\pi\rho; \\
    -\i \nabla_{0} (K^{i} - L^{i}) + e^{ijk} \nabla_{j} (K_{k} -
    L_{k}) = \i \frac{4\pi}{c} j^{i}.
  \end{gathered}
\end{equation}

\subsubsection{Комплексное представление уравнений Максвелла в
  вакууме}
\label{sec:complex:vacuum}

Из соотношений $D^{i} = E^{i}$, $H^{i} = B^{i}$ и
\eqref{eq:complex:K_L} получаем
\begin{equation}
  \label{eq:complex:K_L:vacuum}
  K^{i} = E^{i} + \i B^{i} = F^{i}, \quad L^{i} = 0.
\end{equation}

Тогда уравнения \eqref{eq:complex:maxwell_K_L} будут иметь вид
\begin{equation}
  \label{eq:complex:maxwell_K_L:vacuum}
  \begin{gathered}
    \nabla_{i} F^{i} = 4\pi\rho; \\
    -\i \nabla_{0} F^{i} + e^{ijk} \nabla_{j} F_{k} 
    = \i \frac{4\pi}{c} j^{i}.
  \end{gathered}
\end{equation}

\subsubsection{Комплексное представление уравнений Максвелла в
  однородной изотропной среде}
\label{sec:complex:isotrop}

В однородной изотропной среде справедливы следующие соотношения $D^{i}
= \varepsilon E^{i}$, $\mu H^{i} = B^{i}$, где $\varepsilon$ и
$\mu$~--- диэлектрическая и магнитная проницаемости.

Для упрощения получаемых выражений возможен следующий трюк. В
\eqref{eq:complex:maxwell_K_L:vacuum} делаем формальную замену $c \to
c' = \frac{c}{\sqrt{\varepsilon\mu}}$ (то есть заменяем скорость света
в вакууме на скорость света в среде) и $j^{\alpha} \to
\frac{j^{\alpha}}{\sqrt{\varepsilon}}$.
Тогда получим
\begin{equation}
  \label{eq:complex:maxwell:isotrop}
  \begin{gathered}
    F^{i} = \sqrt{\varepsilon} E^{i} + \i \frac{1}{\sqrt{\mu}} B^{i}, \\
    \nabla_{i} F^{i} = \frac{4\pi}{\sqrt{\varepsilon}}\rho; \\
    e^{ijk} \nabla_{j} F_{k} = \i \frac{4\pi\sqrt{\mu}}{c} j^{i}
    + \i \frac{\sqrt{\varepsilon\mu}}{c} \frac{\partial F^{i}}{\partial t}.
  \end{gathered}
\end{equation}

Данное представление уравнений Максвелла имеет несколько наименований.
В частности, оно известно как представление Римана--Зильберштейна.

\subsection{Импульсное представление уравнений Максвелла}
\label{sec:impuls}

Разложим векторы напряжённости электрического и магнитного полей,
электрической и магнитной индукций в ряд Фурье по волновым векторам
$k^j$, $j$~--- абстрактный индекс (фурье-образы стандартно обозначим
шапочкой):
\begin{equation}
  \label{eq:m:fourie}
  \begin{gathered}
    \Hat{E}^i(t,k_j)=\frac{1}{\sqrt{(2\pi)^3}}\int \d^3 x^j \sqrt{g^{(3)}}
    E^i(t,x^j) \e^{-\i k_jx^j},\\
    \Hat{H}^i(t,k_j)=\frac{1}{\sqrt{(2\pi)^3}}\int \d^3 x^j \sqrt{g^{(3)}}
    H^i(t,x^j) \e^{-\i k_jx^j},\\
    \Hat{B}^i(t,k_j)=\frac{1}{\sqrt{(2\pi)^3}}\int \d^3 x^j \sqrt{g^{(3)}}
    B^i(t,x^j) \e^{-\i k_jx^j},\\
    \Hat{D}^i(t,k_j)=\frac{1}{\sqrt{(2\pi)^3}}\int \d^3 x^j \sqrt{g^{(3)}}
    D^i(t,x^j) \e^{-\i k_jx^j},\\
    \Hat{\rho}(t,k_j)=\frac{1}{\sqrt{(2\pi)^3}}\int \d^3 x^j \sqrt{g^{(3)}}
    \rho(t,x^j) \e^{-\i k_jx^j},\\
    \Hat{\jmath}^i(t,k_j)=\frac{1}{\sqrt{(2\pi)^3}}\int \d^3 x^j \sqrt{g^{(3)}}
    j^i(t,x^j) \e^{-\i k_jx^j}.
  \end{gathered}
\end{equation}

Обратное преобразование:
\begin{equation}
  \label{eq:m:fourie}
  \begin{gathered}
    E^i(t,x^j)=\frac{1}{\sqrt{(2\pi)^3}}\int \d^3 k_j \sqrt{\Hat{g}^{(3)}}
    \Hat{E}^i(t,k_j) \e^{\i k_jx^j},\\
    H^i(t,x^j)=\frac{1}{\sqrt{(2\pi)^3}}\int \d^3 k_j \sqrt{\Hat{g}^{(3)}}
    \Hat{H}^i(t,k_j) \e^{\i k_jx^j},\\
    B^i(t,x^j)=\frac{1}{\sqrt{(2\pi)^3}}\int \d^3 k_j \sqrt{\Hat{g}^{(3)}}
    \Hat{B}^i(t,k_j) \e^{\i k_jx^j},\\
    D^i(t,x^j)=\frac{1}{\sqrt{(2\pi)^3}}\int \d^3 k_j \sqrt{\Hat{g}^{(3)}}
    \Hat{D}^i(t,k_j) \e^{\i k_jx^j}, \\
    \rho(t,x^j)=\frac{1}{\sqrt{(2\pi)^3}}\int \d^3 k_j \sqrt{\Hat{g}^{(3)}}
    \Hat{\rho}(t,k_j) \e^{\i k_jx^j}, \\
    j^i(t,x^j)=\frac{1}{\sqrt{(2\pi)^3}}\int \d^3 k_j \sqrt{\Hat{g}^{(3)}}
    \Hat{\jmath}^i(t,k_j) \e^{\i k_jx^j}.
  \end{gathered}
\end{equation}

Следует заметить, что компоненты векторов $E^i(t,x^j)$ и $\Hat{E}^i(t,k_j)$,
(аналогично: $H^i(t,x^j)$ и $\Hat{H}^i(t,k_j)$, $D^i(t,x^j)$ и $\Hat{D}^i(t,k_j)$,
$B^i(t,x^j)$ и $\Hat{B}^i(t,k_j)$, $j^i(t,x^j)$ и $\Hat{\jmath}^i(t,k_j)$) 
рассматриваются в разных базисах:
\begin{equation}
\begin{gathered}
E^i(t,x^j)=E^{\crd{i}}(t,x^j)\delta^i_{\crd{i}},\\
\Hat{E}^i(t,k_j)=\Hat{E}^{\crd{\Hat{\imath}}}(t,k_j)\delta^i_{\crd{\Hat{\imath}}}, \\
\Hat{g}^{(3)} := \det g_{\crd{\Hat{\imath}} \crd{\Hat{\jmath}}}, \quad 
\d s^2 = g_{\crd{\Hat{\imath}} \crd{\Hat{\jmath}}} \d x^{\crd{\Hat{\imath}}} \d x^{\crd{\Hat{\jmath}}}.
\end{gathered}
\end{equation}
где базис $\delta^i_{\Hat{\crd{\imath}}}$ взят относительно вектора $k_i$.
Для всех $k_i$ определён свой независимый базис. При
выписывании уравнений Максвелла из \eqref{eq:m:fourie} можно работать
не с интегралами, а напрямую с подынтегральными выражениями. Или
воспользоваться формулами для преобразований Фурье:
\begin{equation}
\label{eq:fourier_transforms}
\begin{gathered}
\widehat{(af(x^i)+bg(x^i))} = a \Hat{f}(k_i) + b \Hat{g}(k_i), \quad a,
b = \const, \\
\widehat{\frac{\partial f(x^i) }{\partial x^{\crd{j}}}} = \i
k_{\crd{j}} \Hat{f}(k^i), \\
\widehat{f(x^i) g(x^i)} = \frac{1}{\sqrt{(2\pi)^3}} (\Hat{f} *
\Hat{g}) (k_{i}), \\ 
\shortintertext{где свёртка имеет вид} 
(\Hat{f} *\Hat{g}) (k_{i}) =
\int_{-\infty}^{\infty} \Hat{f}(k_{i} - s_{i}) \Hat{g}(s_{i}) \d^3 s_{i}.
\end{gathered}
\end{equation}

Считая $g = \const$, получим

\begin{equation}
  \label{eq:maxwell:k}
  \begin{gathered}
    \i \frac{1}{\sqrt{g^{(3)}}}\varepsilon^{ijk} k_j E_{k}(t,k_j)=-\frac{1}{c}\partial_t B^i(t, k_j),\\
    \i\frac{1}{\sqrt{g^{(3)}}}\varepsilon^{ijk}k_j H_k(t,k_j)=\frac{1}{c}\partial_t D^i(t,k_j)+\frac{4\pi}{c}j^i(t,k_j),\\
    \i k_i D^i(t, k_j)=4 \pi\rho(t,k_j),\\
    \i k_i B^i(t, k_j)=0.
  \end{gathered}
\end{equation}

  Поскольку результирующие \eqref{eq:maxwell:k} уравнения получаются
  комплексными, то представляется более оправданным использование в
  данном подходе комплексного представления уравнений
  Максвелла~\eqref{eq:complex:eqiv}:
\begin{equation}
  \label{eq:m:fourie:complex}
  \begin{gathered}
    F^i(t,x^j)=\frac{1}{\sqrt{(2\pi)^3}}\int \d^3 k_j \sqrt{\Hat{g}^{(3)}}
    \Hat{F}^i(t,k_j) \e^{\i k_jx^j},\\
    G^i(t,x^j)=\frac{1}{\sqrt{(2\pi)^3}}\int \d^3 k_j \sqrt{\Hat{g}^{(3)}}
    \Hat{G}^i(t,k_j) \e^{\i k_jx^j},\\
    \rho(t,x^j)=\frac{1}{\sqrt{(2\pi)}}\int \d^3 k_j \sqrt{\Hat{g}^{(3)}}
    \Hat{\rho}(t,k_j) \e^{\i k_jx^j}, \\
    j^i(t,x^j)=\frac{1}{\sqrt{(2\pi)^3}}\int \d^3 k_j \sqrt{\Hat{g}^{(3)}}
    \Hat{\jmath}^i(t,k_j) \e^{\i k_jx^j}.
  \end{gathered}
\end{equation}

\textbf{Замечание.} 
  В рамках классической электродинамики разложение векторов $E^j$,
  $H^j$, $B^j$, $D^j$ в ряд Фурье по волновым векторам $k^j$
  соответствует в квантовой механике разложению этих векторов в ряд
  Фурье по импульсам. Поэтому представление \eqref{eq:maxwell:k} можно
  назвать импульсным.

\subsection{Спинорная запись уравнений Максвелла}
\label{sec:spinor}

Тензор электромагнитного поля $F_{\alpha\beta}$ и его компоненты $F_{\crd{\alpha}\crd{\beta}}$, $\crd{\alpha},
\crd{\beta}=\overline{0,3}$, можно рассматривать в спинорной форме
\cite[с. 153]{penrose-rindler-1987} (аналогично и для
\(G_{\alpha\beta}\)): 
\begin{equation}
\label{eq:spinor:F_as_spinor}
\begin{gathered}
F_{\alpha\beta} = F_{AA'BB'}; \\
F_{\crd{\alpha}\,\crd{\beta}}=F_{\crd{A}\,\crd{A}' \crd{B}\,\crd{B}'}
\tensor{g}{_{\crd{\alpha}}}{^{\crd{A}\,\crd{A}'}}\tensor{g}{_{\crd{\beta}}}{^{\crd{B}\,\crd{B}'}}, \\
\crd{A}, \crd{A}', \crd{B}, \crd{B}' =\overline{0,1},\quad \crd{\alpha},
\crd{\beta}=\overline{0,3},
\end{gathered}
\end{equation}
где $\tensor{g}{_{\crd{\alpha}}}{^{\crd{A}\,\crd{A}'}}$,
$\crd{\alpha}=\overline{0,3}$,~--- символы Инфельда--ван дер Вердена,
определяемые в действительном спинорном базисе $\varepsilon_{\crd{A}\,\crd{B}}$
 следующим
образом~\cite[с. 161]{penrose-rindler-1987}:
\begin{equation}
\label{eq:g-iv}
\tensor{g}{_{\crd{\alpha}}}{^{\crd{A}\,\crd{A}'}}:=\tensor{g}{_{\crd{\alpha}}}{^{\alpha}}
\tensor{\varepsilon}{_{A}}{^{\crd{A}}} \tensor{\varepsilon}{_{A'}}{^{\crd{A}'}},\quad 
\tensor{g}{_{\crd{A}\,\crd{A}'}}{^{\crd{\alpha}}}:=\tensor{g}{^{\crd{\alpha}}}{_{\alpha}}
\tensor{\varepsilon}{^{A}}{_{\crd{A}}} \tensor{\varepsilon}{^{A'}}{_{\crd{A}'}},
\end{equation}
\begin{equation}
\label{spin-basis}
\varepsilon_{\crd{A}\,\crd{B}}=\varepsilon_{\crd{A}'\,\crd{B}'}=\begin{pmatrix}
0 & 1\\
-1 & 0
\end{pmatrix},\quad
\tensor{\varepsilon}{_{\crd{A}}}{^A}
\tensor{\varepsilon}{_A}{^{\crd{B}}}= \tensor{\varepsilon}{_{\crd{A}}}{^{\crd{B}}}=\begin{pmatrix}
1 & 0\\
0 & 1
\end{pmatrix}
.
\end{equation}

Запишем уравнения Максвелла через спиноры.

Поскольку тензор \(F_{\alpha\beta}\) действителен и антисимметричен,
то его можно представить в виде
\begin{gather}
  \label{eq:F_spinor}
  F_{\alpha\beta} = \varphi_{AB} \varepsilon_{A'B'} + \varepsilon_{AB}
  \Bar{\varphi}_{A'B'}, \\
  \label{eq:*F^spinor}
  \prescript{*}{}{F}^{\alpha\beta} = - \i \varphi^{AB}\varepsilon^{A'B'} +
  \i \varepsilon^{AB} \Bar{\varphi}^{A'B'}.
\end{gather}
здесь $\varphi_{AB}$~--- спинор электромагнитного поля:
\[
\varphi_{AB} \vcentcolon= \frac{1}{2}\tensor{F}{_{ABC'}}{^{C'}}=\frac{1}{2}F_{AA'BB'}\varepsilon^{A'B'}=\frac{1}{2}F_{\alpha\beta}\varepsilon^{A'B'}. 
\]

Аналогично можно записать
\begin{gather}
  \label{eq:G^spinor}
  G^{\alpha\beta} = \gamma^{AB}\varepsilon^{A'B'} +
  \varepsilon^{AB} \Bar{\gamma}^{A'B'}, \\
  \label{eq:*G_spinor}
  \prescript{*}{}{G}_{\alpha\beta} = - \i \gamma_{AB}\varepsilon_{A'B'} +
  \i \varepsilon_{AB} \Bar{\gamma}_{A'B'}.
\end{gather}

Заменяя в уравнении \eqref{eq:m:tensor} абстрактные индексы $\alpha$
на $AA'$ и $\beta$ на $BB'$, запишем:
\[
\nabla_{AA'} G^{AA' BB'}=\frac{4 \pi}{c}j^{BB'}.
\]

Используя соотношение~\eqref{eq:G^spinor} получим 
\begin{equation}
  \label{eq:spinor:maxwell:G}
  \nabla^{AB'} \gamma^{B}_{A} + \nabla^{BA'} \gamma^{B'}_{A'} =
  \frac{4 \pi}{c}j^{BB'}.
\end{equation}

Аналогично, из \eqref{eq:m:tensor:2:dual} и \eqref{eq:*F^spinor}
получим
\begin{equation}
  \label{eq:spinor:maxwell:*G}
  \nabla^{A'B} \varphi^{A}_{B} - \nabla^{AB'} \Bar{\varphi}^{A'}_{B'} = 0.
\end{equation}

Таким образом полная система уравнений Максвелла в спинорном
представлении имеет вид
\begin{equation}
  \label{eq:spinor:maxwell}
  \begin{gathered}
    \nabla^{A'B} \varphi^{A}_{B} - \nabla^{AB'} \Bar{\varphi}^{A'}_{B'}
    = 0, \\
    \nabla^{AB'} \gamma^{B}_{A} + \nabla^{BA'} \gamma^{B'}_{A'} =
    \frac{4 \pi}{c}j^{BB'}.
  \end{gathered}
\end{equation}

Система уравнений Максвелла в вакууме в спинорной форме запишется
в виде одного уравнения~\cite[с. 385]{penrose-rindler-1987}:
\begin{equation}
  \label{eq:maxwell-spin}
  \nabla^{AB'} \varphi^B_A = \frac{2 \pi}{c}j^{BB'}.
  %
\end{equation}

Выпишем компоненты спинора электромагнитного поля:
\begin{equation}
\begin{gathered}
\varphi_{\crd{A}\,\crd{B}}=\frac{1}{2}F_{\crd{\alpha}\crd{\beta}}\varepsilon^{\crd{A'}\,\crd{B'}} 
\tensor{g}{^{\crd{\alpha}}}{_{\crd{A}\,\crd{A'}}}
\tensor{g}{^{\crd{\beta}}}{_{\crd{B}\,\crd{B'}}},  \\
\crd{A}, \crd{A}', \crd{B}, \crd{B}' =\overline{0,1},\quad \crd{\alpha},
\crd{\beta}=\overline{0,3}.
\end{gathered}
\end{equation}

Используя \eqref{eq:g-iv}, \eqref{spin-basis} и обозначив
$F_{i}=E_{i} - \i B^{i}$, можно
записать~\cite[с. 386]{penrose-rindler-1987}:
\begin{equation}
\begin{gathered}
\varphi_{00}=\frac{1}{2}\left(F_1 - \i F_2\right), \\
\varphi_{01}=\varphi_{10}=-\frac{1}{2} F_3,\\
\varphi_{11}=-\frac{1}{2}\left( F_1 + \i F_2\right).
\end{gathered}
\end{equation}

  \section{Реализация уравнений Максвелла в некоторых системах
    координат}
\label{sec:realisation}

Продемонстрируем реализацию уравнений Максвелла в голономном базисе на
примере часто используемых систем координат: цилиндрической и
сферической. Результат можно сравнить с реализацией в неголономной
системе координат~\cite{batygin:2002}.

\subsection{Уравнения Максвелла в цилиндрической системе координат}

В рамках стандарта ISO~31-11 координаты $(x^{1}, x^{2}, x^{3})$
обозначаются как $(\rho, \varphi, z)$.
Чтобы не возникало коллизий с обозначением плотности заряда $\rho$,
будем использовать обозначения $(r, \varphi, z)$

Закон преобразования координат от декартовых к цилиндрическим:
\begin{equation}
  \begin{cases}
    x=r\cos\varphi, \\
    y=r\sin\varphi, \\
    z=z.
  \end{cases}
\end{equation}

Закон преобразования координат от цилиндрических к декартовым:
\begin{equation}
  \begin{cases} 
    r=\sqrt{x^2+y^2}, \\
    \varphi=\arctg \left(\dfrac{y}{x}\right), \\
    z=z.
  \end{cases}
\end{equation}

Метрический тензор:
\begin{equation}
  g_{\crd{i}\crd{j}} =
  \begin{pmatrix} 
    1 & 0 & 0\\ 
    0 & r^2 & 0\\ 
    0 & 0 & 1 
  \end{pmatrix},\quad 
  g^{\crd{i}\crd{j}} = 
  \begin{pmatrix} 
    1 & 0 & 0\\ 
    0 & 1/r^2 & 0\\ 
    0 & 0 & 1 
  \end{pmatrix}.
\end{equation}

\begin{equation}
  \sqrt{g} = r.
\end{equation}

Коэффициенты Ламе:
\begin{equation}
  h_{1} \equiv h_r=1, \quad h_{2} \equiv h_\varphi=r, \quad h_{3} \equiv h_z=1.
\end{equation}

Соотношение между голономным (тензорным) и неголономным (векторным)
базисами (см.~\eqref{eq:vec_nonholonom} и \eqref{eq:covec_nonholonom}):
\begin{gather}
f^{{r}'} = f^{{r}}, \quad
f^{{\varphi}'} = r f^{{\varphi}}, \quad
f^{{z}'} = f^{{z}}, \\
f_{{r}'} = f_{{r}}, \quad
f_{{\varphi}'} = \frac{1}{r} f_{{\varphi}}, \quad
f_{{z}'} = f_{{z}}.
\end{gather}

Дифференциальные операторы в голономном базисе:

\begin{gather}
  (\Grad f)_i= 
  \frac{\partial f}{\partial r} \delta_i^{{r}}
  + \frac{\partial f}{\partial \varphi} \delta_i^{{\varphi}} 
  + \frac{\partial f}{\partial z} \delta_i^{{z}}; \\
  \Div \vec{f} = 
  \frac{1}{r} \partial_{r}\left(r f^{{r}}\right)
  +  \partial_{\varphi}\left(f^{{\varphi}}\right)
  + \partial_{z}\left(f^{{z}}\right); \\
  \begin{multlined}
    \left(\Rot \vec{f} \right)^{i} = \frac{1}{r}
    \left[ \partial_{\varphi} f_{{z}} - \partial_{z}
      f_{{\varphi}} \right] \delta_{{r}}^{i} + {} \\
    {} +
    \frac{1}{r}
    \left[ \partial_{z} f_{{r}} - \partial_{r} f_{{z}} \right]
    \delta_{{\varphi}}^{i} + \frac{1}{r} \left[ \partial_{r}
      f_{{\varphi}} - \partial_{\varphi} f_{{r}} \right]
    \delta_{{z}}^{i}.
  \end{multlined}
\end{gather}

Дифференциальные операторы в неголономном базисе:

\begin{gather}
  (\Grad f)_i= 
  \frac{\partial f}{\partial r} \delta_i^{{r}'}
  + \frac{1}{r}
  \frac{\partial f}{\partial \varphi} \delta_i^{{\varphi}'} 
  + \frac{\partial f}{\partial z} \delta_i^{{z}'}; \\
  \Div \vec{f} = 
  \frac{1}{r} \partial_{r}\left(r f^{{r}'}\right)
  + \frac{1}{r} \partial_{\varphi}\left(f^{{\varphi}'}\right)
  + \partial_{z}\left(f^{{z}'}\right); \\
  \begin{multlined}
    \left(\Rot \vec{f} \right)^{i} = \frac{1}{r}
    \left[ \partial_{\varphi} f_{{z}'} - r \partial_{z}
     f_{{\varphi}'} \right] \delta_{{r}'}^{i} + {} \\
    {} +
    \left[ \partial_{z} f_{{r}'} - \partial_{r} f_{{z}'} \right]
    \delta_{{\varphi}}^{i} + 
    \frac{1}{r} \left[ \partial_{r}
      (r f_{{\varphi}'}) - \partial_{\varphi} f_{{r}'} \right]
    \delta_{{z}'}^{i}.
  \end{multlined}
\end{gather}

Запишем уравнения Максвелла в цилиндрических координатах $(r,\varphi,
z)$.
\begin{equation}
  \label{eq:maxwell_tc:1}
  \begin{gathered}
    \frac{1}{r}\left[\partial_{\crd{j}}E_{\crd{k}}-\partial_{\crd{k}}E_{\crd{j}}\right]
    = -\frac{1}{c}\partial_t B^{\crd{i}}, \quad \crd{i},\,\crd{j},\,
    \crd{k}=\overline{1,3},\\
    \frac{1}{r}\left[\partial_{\crd{j}}H_{\crd{k}}-\partial_{\crd{k}}H_{\crd{j}}\right]
    = -\frac{1}{c}\partial_t D^{\crd{i}}+\frac{4\pi}{c}j^{\crd{i}}, \quad \crd{i},\,\crd{j},\,
    \crd{k}=\overline{1,3},\\
\frac{1}{r}\partial_{\crd{i}}\left(r
  D^{\crd{i}}\right)= 4\pi \rho, \quad \crd{i}=\overline{1,3},\\
\frac{1}{r}\partial_{\crd{i}}\left(r
  B^{\crd{i}}\right)= 0, \quad \crd{i}=\overline{1,3}.
  \end{gathered}
\end{equation}

После преобразований окончательно получаем:
\begin{gather}
  \label{eq:maxwell_tc:2}
    \frac{1}{r}\left[\partial_{\varphi}E_3
      - \partial_zE_2\right]=-\frac{1}{c}\partial_tB^1,\\
    \frac{1}{r}\left[\partial_{z}E_1
      - \partial_rE_3\right]=-\frac{1}{c}\partial_tB^2,\\
    \frac{1}{r}\left[\partial_{r}E_2
      - \partial_\varphi E_1\right]=-\frac{1}{c}\partial_tB^3,\\
    \frac{1}{r}\left[\partial_{\varphi}H_3
      - \partial_zH_2\right]=-\frac{1}{c}\partial_tD^1+ \frac{4\pi}{c}j^1,\\
    \frac{1}{r}\left[\partial_{z}H_1
      - \partial_rH_3\right]=-\frac{1}{c}\partial_tD^2+ \frac{4\pi}{c}j^2,\\
    \frac{1}{r}\left[\partial_{r}H_2
      - \partial_\varphi H_1\right]=-\frac{1}{c}\partial_tD^3+ \frac{4\pi}{c}j^3,\\
    \frac{1}{r}D^1+\frac{\partial D^1}{\partial r}+\frac{\partial
      D^2}{\partial \varphi}+\frac{\partial D^3}{\partial z}=4 \pi\rho,\\
    \frac{1}{r}B^1+\frac{\partial B^1}{\partial r}+\frac{\partial
      B^2}{\partial \varphi}+\frac{\partial B^3}{\partial z}=0.
\end{gather}

\subsection{Уравнения Максвелла в сферической системе координат}

В рамках стандарта ISO~31-11 координаты $(x^{1}, x^{2}, x^{3})$
обозначаются как $(r, \vartheta, \varphi)$.

Закон преобразования координат от декартовых к сферическим:
\begin{equation}
  \begin{cases} 
    x=r\sin\vartheta\cos\varphi, \\
    y=r\sin\vartheta\sin\varphi, \\ 
    z=r\cos\vartheta. \end{cases}
\end{equation}

Закон преобразования координат от сферических к декартовым:
\begin{equation}
  \begin{cases} 
    r=\sqrt{x^2+y^2+z^2}, \\
    \vartheta=\arccos \left({\dfrac{z}{\sqrt{x^2+y^2+z^2}}}\right) =
    \arctg \left({\dfrac{\sqrt{x^2+y^2}}{z}}\right), \\ 
    \varphi=\arctg\left({\dfrac{y}{x}}\right). 
  \end{cases} 
\end{equation}

Метрический тензор:
\begin{equation}
  g_{\crd{i}\crd{j}} = 
  \begin{pmatrix} 
    1 & 0 & 0\\ 
    0 & r^2 & 0\\ 
    0 & 0 & r^2\sin^2\vartheta 
  \end{pmatrix}, \quad 
  g^{\crd{i}\crd{j}} =
  \begin{pmatrix} 
    1 & 0 & 0\\ 0 & 
    \dfrac{1}{r^2} & 0\\ 0 & 0 &
    \dfrac{1}{r^2\sin^2\vartheta} 
  \end{pmatrix} 
\end{equation}
  
\begin{equation}
  \sqrt{g} = r^2 \sin \vartheta.
\end{equation}

Коэффициенты Ламе:
\begin{equation}
  h_{1} \equiv h_r=1, \quad h_{2} \equiv h_\vartheta=r, \quad h_{3} \equiv h_\varphi=r\sin\vartheta.
\end{equation}

Соотношение между голономным (тензорным) и неголономным (векторным)
базисами (см.~\eqref{eq:vec_nonholonom} и \eqref{eq:covec_nonholonom}):
\begin{gather}
f^{{r}'} = f^{{r}}, \quad
f^{{\vartheta}'} = r f^{{\vartheta}}, \quad
f^{{\varphi}'} = r\sin\vartheta f^{{\varphi}}, \\
f_{{r}'} = f_{{r}}, \quad
f_{{\vartheta}'} = \frac{1}{r} f_{{\vartheta}}, \quad
f_{{\varphi}'} = \frac{1}{r\sin\vartheta} f_{{\varphi}}.
\end{gather}

Дифференциальные операторы в голономном базисе:

\begin{gather}
  (\Grad f)_i= 
  \frac{\partial f}{\partial r} \delta_i^{{r}}
  + \frac{\partial f}{\partial \vartheta} \delta_i^{{\vartheta}} 
  + \frac{\partial f}{\partial \varphi} \delta_i^{{\varphi}}; \\
  \Div \vec{f} = 
  \frac{1}{r^2} \partial_{r}\left(r^2 f^{{r}}\right)
  +  \frac{1}{\sin\vartheta}
  \partial_{\vartheta}\left(\sin\vartheta f^{{\vartheta}}\right)
  + \partial_{\varphi}\left(f^{{\varphi}}\right); \\
  \begin{multlined}
    \left(\Rot \vec{f} \right)^{i} = \frac{1}{r^2 \sin\vartheta}
    \left[ \partial_{\vartheta} f_{{\varphi}} - \partial_{\varphi}
      f_{{\vartheta}} \right] \delta_{{r}}^{i} + {} \\
    {} +
    \frac{1}{r^2 \sin\vartheta}
    \left[ \partial_{\varphi} f_{{r}} - \partial_{r} f_{{\varphi}} \right]
    \delta_{{\vartheta}}^{i} + \frac{1}{r^2 \sin\vartheta} \left[ \partial_{r}
      f_{{\vartheta}} - \partial_{\vartheta} f_{{r}} \right]
    \delta_{{\varphi}}^{i}.
  \end{multlined}
\end{gather}

Дифференциальные операторы в неголономном базисе:

\begin{gather}
  (\Grad f)_i= 
  \frac{\partial f}{\partial r} \delta_i^{{r}'}
  + \frac{1}{r}
  \frac{\partial f}{\partial \vartheta} \delta_i^{{\vartheta}'} 
  + \frac{1}{r\sin\vartheta}
  \frac{\partial f}{\partial \varphi} \delta_i^{{\varphi}'}; \\
  \begin{multlined}
    \Div \vec{f} = 
    \frac{1}{r^2} \partial_{r}\left(r^2 f^{{r}'}\right)
    +  \frac{1}{r\sin\vartheta}
    \partial_{\vartheta}\left(\sin\vartheta
      f^{{\vartheta}'}\right) + {} \\ {}
    + \frac{1}{r\sin\vartheta}
    \partial_{\varphi}\left(f^{{\varphi}'}\right); 
  \end{multlined}
  \\
  \begin{multlined}
    \left(\Rot \vec{f} \right)^{i} = \frac{1}{r \sin\vartheta}
    \left[ \partial_{\vartheta}(\sin\vartheta f_{{\varphi}'}) - \partial_{\varphi}
      f_{{\vartheta}'} \right] \delta_{{r}'}^{i} + {} \\
    {} +
    \frac{1}{r}
    \left[ \frac{1}{\sin\vartheta} \partial_{\varphi} f_{{r}'}
      - \partial_{r} (r f_{{\varphi}'}) \right]
    \delta_{{\vartheta}'}^{i} + \frac{1}{r} \left[ \partial_{r}
     (r f_{{\vartheta}'}) - \partial_{\vartheta} f_{{r}'} \right]
    \delta_{{\varphi}'}^{i}.
  \end{multlined}
\end{gather}

Запишем уравнения Максвелла в сферических координатах
$(r,\vartheta,\varphi)$.
\begin{equation}
  \label{eq:maxwell_sf:1}
  \begin{gathered}
    \frac{1}{r^2\sin\vartheta}\left[\partial_{\crd{j}}E_{\crd{k}}-\partial_{\crd{k}}E_{\crd{j}}\right]
    = -\frac{1}{c}\partial_t B^{\crd{i}}, \quad \crd{i},\,\crd{j},\,
    \crd{k}=\overline{1,3},\\
    \frac{1}{r^2\sin\vartheta}\left[\partial_{\crd{j}}H_{\crd{k}}-\partial_{\crd{k}}H_{\crd{j}}\right]
    = -\frac{1}{c}\partial_t D^{\crd{i}}+\frac{4\pi}{c}j^{\crd{i}}, \quad \crd{i},\,\crd{j},\,
    \crd{k}=\overline{1,3},\\
\frac{1}{r^2\sin\vartheta}\partial_{\crd{i}}\left(r^2\sin\vartheta
  D^{\crd{i}}\right)= 4\pi \rho, \quad \crd{i}=\overline{1,3},\\
\frac{1}{r^2\sin\vartheta}\partial_{\crd{i}}\left(r^2\sin\vartheta
  B^{\crd{i}}\right)= 0, \quad \crd{i}=\overline{1,3}.
  \end{gathered}
\end{equation}

После преобразований окончательно получаем:
\begin{gather}
  \label{eq:maxwell_sf:2}
    \frac{1}{r^2\sin\vartheta}\left[\partial_{\vartheta}E_3
      - \partial_\varphi E_2\right]=-\frac{1}{c}\partial_tB^1,\\
    \frac{1}{r^2\sin\vartheta}\left[\partial_{\varphi}E_1
      - \partial_rE_3\right]=-\frac{1}{c}\partial_tB^2,\\
    \frac{1}{r^2\sin\vartheta}\left[\partial_{r}E_2
      - \partial_\vartheta E_1\right]=-\frac{1}{c}\partial_tB^3,\\
    \frac{1}{r^2\sin\vartheta}\left[\partial_{\vartheta}H_3
      - \partial_\varphi H_2\right]=-\frac{1}{c}\partial_tD^1+ \frac{4\pi}{c}j^1,\\
    \frac{1}{r^2\sin\vartheta}\left[\partial_{\varphi}H_1
      - \partial_rH_3\right]=-\frac{1}{c}\partial_tD^2+ \frac{4\pi}{c}j^2,\\
    \frac{1}{r^2\sin\vartheta}\left[\partial_{r}H_2
      - \partial_\vartheta H_1\right]=-\frac{1}{c}\partial_tD^3+ \frac{4\pi}{c}j^3,\\
    \frac{2}{r}D^1+\partial_rD^1+\ctg \vartheta D^2+\partial_\vartheta
    D^2+\partial_\varphi D^3=4\pi\rho,\\
    \frac{2}{r}B^1+\partial_rB^1+\ctg \vartheta B^2+\partial_\vartheta
    B^2+\partial_\varphi B^3=0.
\end{gather}

\section{Заключение}

Сформулируем основные выводы и результаты, полученные в работе:
\begin{enumerate}
\item Показана связь между тензорным и векторным формализмами.
\item Выписано ковариантное координатное представление
  дифференциальных операторов для голономных систем координат.
\item Продемонстрировано применение тензорного формализма для разных
  форм записи уравнений Максвелла.
\item Выписаны уравнения Максвелла в ковариантной бескоординатной и
  ковариантной координатной формах.
\item Показано совпадение результатов, полученных с помощью тензорного
  и векторного формализмов на примере цилиндрической и сферической
  систем координат.
\item Применение для уравнений Максвелла вместо векторного
  формализма тензорного позволяет упростить математические выкладки, в
  частности при работе с недекартовыми системами координат.

  Упрощение записи уравнений и промежуточных расчётов в недекартовых
  системах координат возможно а счёт использования хорошо
  разработанного формализма тензорного анализа. Перевод же в уравнений
  или результатов векторный формализм при необходимости можно
  осуществлять на заключительном этапе.
\end{enumerate}

\bibliographystyle{ugost2008}%

\bibliography{bib/ref}